\documentclass{jpp}

\expandafter\let\csname equation*\endcsname\relax
\expandafter\let\csname endequation*\endcsname\relax

\usepackage[utf8]{inputenc}
\usepackage[T1]{fontenc}
\usepackage{newtxtext,newtxmath}
\usepackage{indentfirst}

\usepackage{graphics}      
\usepackage{graphicx}      
\usepackage{longtable}     
\usepackage{url}           
\usepackage{amsmath, amssymb, amsbsy}
\usepackage{natbib}
\usepackage{bm}            
\usepackage{dcolumn}       
\usepackage{xcolor}		   
\usepackage[normalem]{ulem} 
\usepackage[section]{placeins} 

\usepackage{lipsum}
\usepackage{apptools}
\usepackage{physics} 
\usepackage{upgreek}
\usepackage[version=4]{mhchem} 
\usepackage[linesnumbered,ruled,vlined, figure]{algorithm2e}
\usepackage[superscript,biblabel]{cite}
\usepackage[font={small}]{caption}
\usepackage[colorlinks=true,allcolors=blue,breaklinks=true,pageanchor=false]{hyperref} 

\usepackage{dcolumn}       
\usepackage{xcolor}		   
\usepackage[normalem]{ulem} 
\usepackage{appendix}
\usepackage{lipsum}
\usepackage{upgreek}

\usepackage[colorlinks=true,allcolors=blue,breaklinks=true,pageanchor=false]{hyperref}

\DeclareMathOperator{\sign}{sgn}

\shorttitle{A Ritz variational principle for local collisionless gyrokinetic instabilities}
\shortauthor{C. D. Stephens, P.-Y. Li}

\title{A Ritz variational principle for local collisionless gyrokinetic instabilities}
\author{C. D. Stephens\aff{1, \corresp{\email{cole.stephens@austin.utexas.edu}}}, P.-Y. Li\aff{1}}
\affiliation{
\aff{1} Institute of Fusion Studies, University of Texas at Austin, TX 78712-1192, United States of America}

\begin{document}
\maketitle

\begin{abstract}
    Turbulence driven by gyrokinetic instabilities is largely responsible for transport in magnetic fusion devices. To estimate this turbulent transport, integrated modeling codes often use mixing length estimates in conjunction with reduced models of the linearized gyrokinetic equation. One common method of formulating and solving the linearized gyrokinetic eigenvalue problem equation uses a Ritz variational principle, particularly in the local collisionless limit. However, the variational principle as typically stated in the literature is mathematically incorrect. In this work, we derive a mathematically correct form of the variational principle that applies to local linear collisionless gyrokinetics in general geometry with electromagnetic effects. We also explicitly derive a weak form of the gyrokinetic field equations suitable for numerical applications. 
\end{abstract}

\section{Introduction}
\label{sec:introduction}
Gyrokinetic modeling is the most advanced and first principles-based method for predicting turbulent transport driven by microinstabilities in magnetic fusion devices \citep{brizard2007gyrokinetic, cary2009guiding}. Although nonlinear simulations can accurately predict particle, momentum, and heat transport \citep{bourdelle2016}, they are often too computationally expensive for integrated modeling purposes. The cost of any given nonlinear simulation is typically on the order of $10^4$ CPUh to $10^5$ CPUh at a single radial point \citep{citrin2017}. Integrated modeling frameworks, meanwhile, require thousands of flux calculations for every second of a plasma discharge in a large magnetic confinement fusion device. Therefore, it is computationally advantageous to approximate the turbulent transport from local linear simulations and make use of a mixing length estimate \citep{casati2009, bourdelle2015, staebler2024}. Quasilinear models, such as TGLF and QuaLiKiz, heavily approximate the linearized gyrokinetic equations to further reduce the computational cost of any given simulation \citep{waltz1997, staebler2007, citrin2017, stephens2021}. These models typically formulate the gyrokinetic equations as an eigenvalue problem instead of an initial value problem. This approach allows for systematic reductions of the resulting equations. Moreover, eigenvalue codes are often necessary since accurate quasilinear models can require the solutions of multiple eigenmodes rather than just that of the most dominant instability \citep{pueschel2016}. In contrast, initial value codes can only obtain the dominant instability. 

One popular way of formulating the gyrokinetic eigenvalue problem is with the use of a Ritz variational principle, which shares some similarities with the action principle. In previous work, the variational principle has also been used to analyze other modes in plasma physics, such as drift-tearing modes and drift waves \citep{ross1978, hazeltine1978, mahajan1979}. Some example applications of this principle to gyrokinetics can be found in \citet{garbet1990}, \citet{bourdelle2002}, \citet{citrin2017}, \citet{hamed2019}, \citet{kotschenreuther2024}, and \citet{morren2024}. The idea is as follows: suppose we have an eigenvalue problem of the form $\mathcal{L}(\omega) \chi = 0$, where $\omega = \omega_r + i \gamma$ is a complex eigenvalue where $\omega_r$ and $\gamma$ are real, $\mathcal{L}$ is an integro-differential operator, and $\chi$ is a complex-valued field that represents the eigenmode. One then left-multiplies by $\chi^{\ast}$ and integrates over the entire domain with a suitable weighting function to obtain a weak form of the problem. For gyrokinetics, one usually solves for the distribution function in terms of the perturbed electromagnetic potentials and fields and then substitutes it into the gyrokinetic field equations. In practice, $\chi$ is either guessed from other methods or taken from a gyrokinetic simulation, though in principle one can solve for $\chi$ self-consistently with the use of trial functions. The mode is considered unstable if the growth rate $\gamma$ is positive; otherwise it is considered stable. 

Unfortunately, for the gyrokinetic field equations, the above method is mathematically incorrect. Although the above specific formulation of the principle was likely taken either from Ritz methods in quantum mechanics or the action principle in field theory, a closer examination of nonlinear eigenvalue problems reveals that it is simply incorrect in gyrokinetics. The fatal flaw is that the gyrokinetic field equations are a nonlinear eigenvalue problem \citep{voss2013}, while the above method was formulated for linear Hermitian eigenvalue problems. The gyrokinetic field equations are not linear in the eigenvalue $\omega$ due to the Landau resonance. In addition, gyrokinetic field equations are not Hermitian because one needs to analytically continue the integral equations with the use of a Landau contour for damped eigenmodes, i.e. modes where the eigenvalues have negative imaginary part. (Note that in nonlinear eigenvalue problems, we take ``Hermitian'' to mean that $\hat{L}(\omega)^{\dagger} = \hat{L}(\omega^{\ast})$, where $\dagger$ denotes the conjugate transpose and the asterisk denotes the scalar complex conjugate.) 

This work has two key goals. The first is to derive and state a mathematically rigorous variational principle for the collisionless local linear gyrokinetic field equations. In general, the rigorous principle for nonlinear non-Hermitian eigenvalue matrix problems requires the use of a left-eigenfunction that solves the adjoint problem \citep{voss2013}. Fortunately, although the gyrokinetic field equations are not Hermitian, they are indeed complex symmetric, meaning that $\hat{L}(\omega) = \hat{L}^T (\omega)$, where $T$ denotes the ordinary transpose. Remarkably, we can then rescue the above incorrect variational principle with a slight modification. Instead of left-multiplying by $\chi^{\ast}$, one merely needs to left-multiply by $\chi$ without taking the complex conjugate and then proceed as usual. We present a heuristic proof of this complex symmetry for the collisionless gyrokinetic field equations, as well as a brief overview of nonlinear eigenvalue problems in Sec.~\ref{sec:variational_intro}. 

The second goal is to derive the specific equations to be used in the variational principle. We rigorously prove the complex symmetry by inspection and compactly represent a weak form for the linearized field equations in general geometry and with electromagnetic effects in the local collisionless limit. Not only is the variational principle now formulated correctly, but it also allows for generic trial functions with linear and nonlinear undetermined coefficients. One can then use a finite element method with adaptive mesh refinement, for example, to solve the gyrokinetic field equations, although we do not do so here. In deriving the equations, we also obtain physical insight into the system via the connection between the gyrokinetic field equations and the use of action angle variables in guiding center orbit theory \citep{stephens2020}. A weak form of the field equations is rigorously derived in Sec.~\ref{sec:solutions}, and we also present a brief analysis of guiding center orbits in Appendix~\ref{sec:guiding} with special attention paid to tokamak and stellarator geometries. Also, although we include the parallel perturbed magnetic field $\delta B_{\parallel}$ in Sec.~\ref{sec:variational_intro}, for the sake of brevity we do not include it in Sec.~\ref{sec:solutions}. Extending the derived equations to include $\delta B_{\parallel}$ is straightforward. 

We note that the general methods and integral transforms used to derive the equations are not new. Many of the techniques can be found in \citet{rewoldt1982} and \citet{garbet1990}. We consider our work an extension of this previous work. Firstly, we have generalized their approach to account for (periodized) stellarator geometry, when their methods were originally designed for tokamak geometry. Secondly, the integral transforms in \citet{rewoldt1982} have been modified such that the dispersion relation resembles the form presented in \citet{garbet1990}. In particular, the drift resonance for passing particles is now present in the resonant denominator with a continuum integral, which should be more convenient for numerical applications. Lastly, while both \citet{rewoldt1982} and \citet{garbet1990} only allowed for the variation of linear coefficients, our formulation allows for both linear and nonlinear varying parameters with any suitable set of trial functions. 

After we discuss the results in Sec.~\ref{sec:conclusions}, we also present a similar analysis for the linearized gyrokinetic equation proper in Appendix~\ref{sec:adjoint} using the perturbed distribution function. The adjoint state method has been recently applied to the linearized gyrokinetic equations \citep{acton2024}, where an adjoint equation is derived and a corresponding adjoint solution is obtained via simulation. We briefly derive a connection between the right eigenfunction and the left eigenfunction for the linearized gyrokinetic equation; this may be useful when applying adjoint state methods or if one wishes to use the distribution function to formulate a variational principle. 

\section{The Variational Principle for Nonlinear Eigenvalue Problems}
\label{sec:variational_intro}

\subsection{Linear Hermitian Eigenvalue Problems}
To begin, we review the variational principle for linear Hermitian eigenvalue problems. Since this method is often used in introductory quantum mechanics textbooks, we shall use it to build intuition \citep{griffiths2018, sakurai2011}.

Let $H$ be an $n\times n$ Hermitian matrix (where $H^\dagger = H$). The linear Hermitian eigenvalue problem is to find $\lambda$ such that
\begin{equation}
\label{eq:hermitian_right}
    H \vb{x} = \lambda \vb{x}, 
\end{equation}
has a nontrivial solution $\vb{x}$. This problem is linear because it depends only linearly on $\lambda$. The adjoint problem is
\begin{equation}
\label{eq:hermitian_left}
    \vb{y}^{\dagger} H = \lambda \vb{y}^{\dagger}. 
\end{equation}
A non-trivial solution $\hat{\vb{x}}$ of Eq.~\ref{eq:hermitian_right} is a right eigenvector of $H$, while a non-trivial solution $\hat{\vb{y}}$ of Eq.~\ref{eq:hermitian_left} is a left eigenvector of $H$. We call a solution pair $(\hat{\lambda}, \hat{\vb{x}})$ of Eq.~\ref{eq:hermitian_right} an eigenpair. Meanwhile, if $\hat{\vb{y}}$ and $\hat{\vb{x}}$ share an eigenvalue $\hat{\lambda}$, then $(\hat{\vb{y}}, \hat{\lambda}, \hat{\vb{x}})$ is an eigentriplet. Because $H$ is Hermitian, we know that the eigenvalues are real and that the left and right eigenvectors corresponding to the same eigenvalue coincide. Therefore, it suffices only to consider Eq.~\ref{eq:hermitian_right}. 

We now describe the variational method for Hermitian linear eigenvalue problems. First, we define the Rayleigh-Ritz quotient
\begin{equation}
    R(\vb{x}) = \frac{\vb{x}^\dagger H \vb{x}}{\vb{x}^\dagger \vb{x}}. 
\end{equation}
The variational theorem tells us that
\begin{equation}
    \lambda_{\text{min}} \le R(\vb{x}) \le \lambda_{\text{max}}.
\end{equation}
Moreover, if $(\hat{\lambda}, \hat{\vb{x}})$ is an eigenpair, then $R(\vb{x})$ is stationary at $\vb{x} = \hat{\vb{x}}$. 

The variational principle is commonly used in quantum mechanics, where the variational principle can be extended for infinite-dimensional systems. Assume we are given a 1-dimensional Hamiltonian operator $\hat{H}$ and complex-valued trial function $\psi(x; \alpha)$ that is a function of position $x$ and parameterized by complex numbers $\alpha$. We can then estimate the ground state by minimizing
\begin{equation}
    R(\alpha) = \frac{\int_{-\infty}^{\infty} \dd{x} \psi(x; \alpha)^\ast \hat{H} \psi(x; \alpha)}{\int_{-\infty}^{\infty} \dd{x} \psi(x; \alpha)^\ast \psi(x; \alpha)}. 
\end{equation}
In this case, if we find a solution $\hat{\alpha}$ such that
\begin{equation}
    \frac{\partial R}{\partial \alpha} = 0, 
\end{equation}
and such that $R(\alpha)$ is minimized, then the estimates for the ground state eigenfunction $\psi_0(x)$ and eigenvalue $\lambda_0$ are
\begin{align}
    \psi_0(x) &\approx \psi(x, \hat{\alpha}),\\
    \lambda_0(x) &\approx R(\alpha_0). 
\end{align}

\subsection{Nonlinear Eigenvalue Problems}
We now extend the discussion to nonlinear eigenvalue problems. A more thorough introduction to the subject can be found in \citet{voss2013}. Let $T(\lambda)$ be an analytic $n\times n$ matrix function that depends on the complex parameter $\lambda$ (in general, nonlinearly). The nonlinear eigenvalue problem is to find a non-trivial eigenpair $(\hat{\lambda}, \hat{\vb{x}})$ that solves the equation
\begin{equation}
    T(\lambda) \vb{x} = \vb{0}. 
\end{equation}
Likewise, the adjoint problem is given by
\begin{equation}
    \vb{y}^{\dagger} T(\lambda) = \vb{0}. 
\end{equation}
We note that the linear Hermitian eigenvalue problem is simply a special case of the nonlinear eigenvalue problem with $T(\lambda) = H - \lambda I$. However, nonlinear eigenvalue problems have several peculiar properties in comparison to linear eigenvalue problems. In particular:
\begin{enumerate}
  \item Eigenvectors corresponding to distinct eigenvalues are not necessarily linearly independent. (They are in linear eigenvalue problems.)
  \item Left and right eigenvectors are not necessarily orthogonal to one another. (They are in linear eigenvalue problems.) 
  \item Even in finite dimensions, there may exist an infinite number of distinct eigenvalues. (There exist only a finite number of distinct eigenvalues in finite-dimensional linear eigenvalue problems.) 
\end{enumerate}
Moreover, a variational characterization of the nonlinear eigenvalue problem is more complicated than that of the linear Hermitian problem. Because the eigenvalues are in general complex, a min-max characterization may not exist. We can, however, still construct a variational principle. Consider the Rayleigh quotient defined by
\begin{equation}
    R(\vb{y}, \lambda, \vb{x}) = \lambda - \frac{\vb{y}^{\dagger} T(\lambda) \vb{x}}{\vb{y}^\dagger T'(\lambda) \vb{x}},
\end{equation}
where $T'(\lambda)$ is the derivative of $T(\lambda)$ with respect to $\lambda$. If $(\hat{\vb{y}}, \hat{\lambda}, \hat{\vb{x}})$ is an eigentriplet and $\hat{\vb{y}}^\dagger T'(\lambda) \hat{\vb{x}} \neq 0 $, then $R$ is stationary at $(\hat{\vb{y}}, \hat{\lambda}, \hat{\vb{x}})$. 

Unlike the linear Hermitian eigenvalue problem, the nonlinear eigenvalue problem uses a Rayleigh quotient that involves both the left and the right eigenvector. This is because, in general, the left and right eigenvectors no longer coincide. Fortunately, a simplification is possible for specific matrices. First, we say $T(\lambda)$ is complex symmetric if
\begin{equation}
    T(\lambda) = T(\lambda)^T
\end{equation}
for every $\lambda$ in our domain of interest. It then follows that if $\hat{\lambda}, \hat{\vb{x}}$ is an eigenpair, then $(\hat{\bar{\vb{x}}}, \hat{\lambda}, \hat{\vb{x}})$ is an eigentriplet, where $\bar{\vb{x}}$ denotes the complex conjugate of $\vb{x}$ without taking the transpose. In this case, we can simplify the Rayleigh quotient to
\begin{equation}
    R(\lambda, \vb{x}) = \lambda - \frac{\vb{x}^T T(\lambda) \vb{x}}{\vb{x}^T T'(\lambda) \vb{x}}. 
\end{equation}
Unlike the Hermitian linear problem, we now consider only the \textit{transpose} of $\vb{x}$ and not the conjugate transpose. We are therefore making use of a null product since $\vb{x}^T \vb{x} = 0$ has non-trivial solutions. We note that this is commonly used in non-Hermitian quantum mechanics, where it is called the c-product \citep{moiseyev2011}.

To treat infinite-dimensional problems on the real line, we first define an inner product. Given functions $f(x)$ and $g(x)$, we define the inner product $\left<g, f \right>$ as
\begin{equation}
    \left<g, f \right> = \int_{-\infty}^{\infty} \dd{x} w(x) g(x)^{\ast} f(x),
\end{equation}
where $w(x)$ is a positive definite weighting function. We then say that the operator $\hat{T}(\lambda)$ is complex symmetric if
\begin{equation}
    \left<g, \hat{T}(\lambda) f\right> = \left<f^\ast, \hat{T}(\lambda) g^{\ast}\right>
\end{equation}
for all complex $\lambda$ in our domain of interest. Rather than computing the full Rayleigh quotient, the variational principle can be stated more simply. Let $\psi(x; \alpha)$ be a complex-valued function parameterized by complex numbers $\alpha$. We define the functional $p$ as
\begin{equation}
    p(\lambda, \alpha) = \left<\psi^\ast, \hat{L}(\lambda) \psi\right> = \int_{-\infty}^{\infty} \dd{x} w(x) \psi(x; \alpha) \hat{L}(\lambda) \psi(x; \alpha). 
\end{equation}
Note the absence of an explicit complex conjugate inside the integral. We can obtain estimate right eigenfunctions $\psi$, left eigenfunctions $\psi^\ast$, and eigenvalues $\lambda$ by simultaneously solving
\begin{align}
    p &= 0, \\
    \frac{\partial p}{\partial \alpha} & = 0 ,\\
     \frac{\partial p}{\partial \lambda} & \neq 0. 
\end{align}
These are the relevant general equations to be solved. We now seek to prove that the gyrokinetic field equations are indeed complex symmetric. 

\section{Complex Symmetry of the Gyrokinetic Field Equations}
When writing down the gyrokinetic equations, we work in the ballooning representation where the perturbed fields $(\delta \phi, \delta A_{\parallel}, \delta B_{\parallel})$ are functions of the field line variable $l$. The fields correspond to the electrostatic potential, the parallel vector potential, and the parallel magnetic field respectively. The field equations can be obtained by substituting the perturbed distribution function from the linearized gyrokinetic equation into the quasineutrality equation and Ampere's law. We shall do this explicitly in Sec.~\ref{sec:solutions}, but for now, we simply write the result:
\begin{align}
    \sum_s \frac{e_s^2 n_{0 s}}{T_{0 s}} \delta \phi - \int \dd[3]{v} \frac{ e_s^2 F_{0 s}}{T_{0 s}} J_{0 s} \frac{\omega - n \omega_{\ast s}}{\omega - v_{\parallel} k_{\parallel} - n \omega_{d s}} \left(J_{0 s} \delta \phi - J_{0 s} v_{\parallel} \delta A_{\parallel} + \frac{2 J_{1s}}{k_{\perp} \rho_s} \frac{\mu}{e_s} \delta B_{\parallel} \right) = 0,\\
    \frac{k_{\perp}^2}{\mu_0} \delta A_{\parallel} - \sum_s \int \dd[3]{v} \frac{e_s^2 F_{0 s}}{T_{0 s}}  v_{\parallel} J_{0 s} \frac{\omega - n \omega_{\ast s}}{\omega - v_{\parallel} k_{\parallel} - n \omega_{d s}}\left(J_{0 s} \delta \phi - J_{0 s} v_{\parallel} \delta A_{\parallel} + \frac{2 J_{1s} }{k_{\perp} \rho_s} \frac{\mu}{e_s} \delta B_{\parallel} \right) = 0,\\
    - \frac{\delta B_{\parallel} }{\mu_0} - \sum_s \int \dd[3]{v} e\frac{e_s F_{0 s}}{T_{0 s}} \frac{2 J_{1s} }{k_{\perp} \rho_s} \mu \frac{\omega - n \omega_{\ast s}}{\omega - v_{\parallel} k_{\parallel} - n \omega_{d s}} \left(J_{0 s} \delta \phi - J_{0 s} v_{\parallel} \delta A_{\parallel} + \frac{2 J_{1s}}{k_{\perp} \rho_s} \frac{\mu}{e_s} \delta B_{\parallel} \right) = 0. 
\end{align}
All species-specific quantities are labeled with the subscript $s$. The background quantities are the density $n_0$, the temperature $T$, the Maxwellian $F$, and the background magnetic field $\vb{B}$. The constants $e$ and $m$ represent the charge and mass of each species. Next, $v_\parallel$ and $\mu = m v_{\perp} / e B$ are the velocity parallel to $\vb{B}$ and the magnetic moment, respectively, where $v_{\perp}$ is the speed perpendicular to the magnetic field.  The frequencies present are the eigenvalue $\omega$, the diamagnetic drift frequency $\omega_{\ast}$, and the drift frequency $\omega_{d}$; meanwhile, $n$ is the toroidal mode number. The wavenumber has been separated into a perpendicular part $k_{\perp}$ and a parallel part $k_{\parallel}$. The first is simply a function of $l$ in ballooning space, whereas the second is a differential operator. The Bessel functions $J_0$ and $J_1$ represent the effect of gyroaveraging and are evaluated at $k_{\perp} \rho$, where $\rho = m v_{\perp} / e B$ is the gyroradius. Lastly, $\mu_0$ is the magnetic constant. It is to be understood that the inverse of a differential operator is an integral operator: this operator is written explicitly in Sec.~\ref{sec:solutions}. 

To simplify notation, represent the three fields with the normalized column vector $\bm{\chi} = (e \delta \phi / T_r, \delta A_{\parallel} / \rho_r B_r, \delta B_{\parallel} / B_r)$. The fields are subject to the boundary condition that $\bm{\chi} \to \vb{0}$ as $\left|l\right| \to \infty$. Here, $n_r, T_r, \rho_r, B_r$ are arbitrary reference density, temperature, gyroradius, and magnetic field, respectively. It is also useful to define a reference sound speed $c_r = \sqrt{m_r / T_r}$ and plasma beta $\beta_r = 2 \mu_0 n_r T_r / B_r^2$, where $m_r$ is a reference mass. We define our inner product by the field line average: 
\begin{equation}
    \left<\bm{\chi}_2, \bm{\chi}_1 \right> = \int_{-\infty}^{\infty} \frac{\dd{l}}{B}  \left(\frac{e^2}{T_r^2}\delta \phi_2^\ast \delta \phi_1 + \frac{1}{\rho_r^2 B_r^2} \delta A_{2\parallel}^\ast \delta A_{1\parallel} + \frac{1}{B_r^2} \delta B_{2\parallel}^\ast \delta B_{1\parallel}\right). 
\end{equation}
The field equations can then be expressed in a more convenient matrix form:
\begin{equation}
    \hat{L} \bm{\chi} = \vb{0},
\end{equation}
where
\begin{equation}
    \hat{L} = \begin{pmatrix}
        \hat{L}_{\phi \phi} &  \hat{L}_{\phi A} & \hat{L}_{\phi B} \\
        \hat{L}_{A \phi} & \hat{L}_{A A} & \hat{L}_{A B} \\
        \hat{L}_{B \phi} & \hat{L}_{B A} & \hat{L}_{B B}
    \end{pmatrix}
\end{equation}
and
\begin{align}
    \hat{L}_{\phi \phi} &= \sum_s \frac{Z_s^2 n_{0 s} T_{r}}{n_{r} T_{0 s}} - \int \dd[3]{v} \frac{Z_s^2 F_{0 s} T_r}{n_r T_{0 s}}  J_{0 s} \frac{\omega - n \omega_{\ast s}}{\omega - v_{\parallel} k_{\parallel} - n \omega_{d s}} J_{0 s}, \\
    \hat{L}_{\phi A} & = \sum_s \int \dd[3]{v} \frac{Z_s^2 F_{0 s} T_r}{n_r T_{0 s} }  J_{0 s} \frac{\omega - n \omega_{\ast s}}{\omega - v_{\parallel} k_{\parallel} - n \omega_{d s}} J_{0 s} \frac{v_{\parallel}}{c_r}, \\
    \hat{L}_{\phi B } & = - \sum_s \int \dd[3]{v} \frac{Z_s F_{0 s} T_r}{n_r T_{0 s} }  J_{0 s} \frac{\omega - n \omega_{\ast s}}{\omega - v_{\parallel} k_{\parallel} - n \omega_{d s}} \frac{2 J_{1s}}{k_\perp \rho_s} \frac{\mu B_r}{T_r}, \\
    \hat{L}_{A \phi} & = - \sum_s \int \dd[3]{v} \frac{Z_s^2 F_{0 s} T_r}{n_r T_{0 s}  }\frac{v_{\parallel}}{c_r} J_{0 s} \frac{\omega - n \omega_{\ast s}}{\omega - v_{\parallel} k_{\parallel} - n \omega_{d s}} J_{0 s}, \\
     \hat{L}_{A A} & = \frac{2 k_{\perp}^2}{\beta_r} + \sum_s \int \dd[3]{v} \frac{Z_s^2  F_{0 s} T_r}{n_r T_{0 s}} \frac{v_{\parallel}}{c_r} J_{0 s} \frac{\omega - n \omega_{\ast s}}{\omega - v_{\parallel} k_{\parallel} - n \omega_{d s}} J_{0 s} \frac{v_{\parallel}}{c_r} , \\
    \hat{L}_{A B } & = - \sum_s \int \dd[3]{v} \frac{Z_s F_{0 s} T_r}{n_r T_{0 s}} \frac{v_{\parallel}}{c_r} J_{0 s} \frac{\omega - n \omega_{\ast s}}{\omega - v_{\parallel} k_{\parallel} - n \omega_{d s}} \frac{2 J_{1s}}{k_{\perp} \rho_s} \frac{\mu B_r}{T_r}, \\
    \hat{L}_{B \phi} & =- \sum_s \int \dd[3]{v} \frac{Z_s F_{0 s} T_r}{n_r T_{0 s} } \frac{2 J_{1s}}{k_\perp \rho_s} \frac{\mu B_r}{T_r}   \frac{\omega - n \omega_{\ast s}}{\omega - v_{\parallel} k_{\parallel} - n \omega_{d s}} J_{0 s}, \\
     \hat{L}_{B A} & = \sum_s \int \dd[3]{v} \frac{Z_s F_{0 s} T_r}{n_r T_{0 s}}  \frac{2 J_{1s}}{k_{\perp} \rho_s} \frac{\mu B_r}{T_r} \frac{\omega - n \omega_{\ast s}}{\omega - v_{\parallel} k_{\parallel} - n \omega_{d s}} \frac{v_{\parallel}}{c_r} J_{0 s} , \\
    \hat{L}_{B B } & = - \frac{2}{\beta_r} - \sum_s \int \dd[3]{v} \frac{F_{0 s} T_r} {n_r T_{0 s}} \frac{2 J_{1s}}{k_\perp \rho_s} \frac{\mu B_r}{T_r} \frac{\omega - n \omega_{\ast s}}{\omega - v_{\parallel} k_{\parallel} - n \omega_{d s}} \frac{2 J_{1s}}{k_\perp \rho_s} \frac{\mu B_r}{T_r}. \\
\end{align}
This block matrix form will be convenient for proving the complex symmetry of the gyroinetic field equations. 

To prove complex symmetry, we must first examine the operator $v_{\parallel} k_{\parallel}$. The operator $k_\parallel$ is defined such the energy and magnetic moment are held constant when taking the partial derivative with respect to the field line variable. Therefore, $v_{\parallel}$ and $k_{\parallel}$ do not commute, so the whole operator must be examined together. However, the operator $v_{\parallel} k_{\parallel}$ is effectively antisymmetric when the integrals over velocity space and along the field line are taken into account. To see this, first take $\sigma$ to represent the sign of $v_{\parallel}$ and define the energy $\varepsilon$ as
\begin{equation}
    \varepsilon = \frac{1}{2} m v_{\parallel}^2 + \mu B.
\end{equation}
Recall that
\begin{equation}
    \dd[3]{v} = \sum_{\sigma = \pm 1} \frac{2 \pi B}{m_s^2 \left|v_{\parallel}\right|} \dd{\varepsilon} \dd{\mu}. 
\end{equation}
Let $f$ and $g$ be arbitrary functions of $(l, \varepsilon, \mu, \sigma)$ that decay away at infinity. Then,
\begin{equation}
\begin{split}
     \sum_{\sigma = \pm 1} \int \frac{\dd{l}}{B} \frac{2 \pi B}{m_s^2 \left|v_{\parallel}\right|} \dd{\varepsilon} \dd{\mu} g^\ast v_{\parallel} k_{\parallel} f & = \sum_{\sigma = \pm 1} \int \dd{l} \frac{2 \pi}{m_s^2} \dd{\varepsilon} \dd{\mu} g^\ast \sigma k_{\parallel} f 
     \\ & = \sum_{\sigma = \pm 1} \int \dd{l} \frac{2 \pi}{m_s^2} \dd{\varepsilon} \dd{\mu} f (-\sigma k_{\parallel}) g^\ast 
     \\ & = \sum_{\sigma = \pm 1} \int \frac{\dd{l}}{B} \frac{2 \pi B}{m_s^2 \left|v_{\parallel}\right|} \dd{\varepsilon} \dd{\mu} f (-v_{\parallel} k_{\parallel}) g^{\ast},
    \end{split}
\end{equation}
where we integrated by parts and set the boundary terms to zero. This proves that $v_{\parallel} k_{\parallel}$ is anti-symmetric. Note that one can bring $v_{\parallel} k_{\parallel}$ under the complex conjugate to show that it is also Hermitian, as expected. 

To show that $\hat{L}$ is complex symmetric under our inner product, we can take advantage of the block matrix form:
\begin{equation}
    \hat{L}^T = \begin{pmatrix}
        \hat{L}_{\phi \phi}^T &  \hat{L}_{A \phi}^T & \hat{L}_{B \phi}^T \\
        \hat{L}_{\phi A}^T & \hat{L}_{A A}^T & \hat{L}_{B A}^T \\
        \hat{L}_{\phi B}^T & \hat{L}_{A B }^T & \hat{L}_{B B}^T 
    \end{pmatrix}. 
\end{equation}
We can check this term by term as follows. First, we note that when taking the transpose, we must reverse the order of all operators: 
\begin{equation}
    \left(\int \dd[3]{v} A B C \right)^T = \int \dd[3]{v} C^T B^T A^T. 
\end{equation}
We also note that any function that is independent $l$ commutes with $k_{\parallel}$ and all other functions of $l$. Next, the transpose of the resonant denominator can be computed by simply substituting $v_{\parallel} k_{\parallel} \to - v_{\parallel} k_{\parallel}$. Lastly, since we are integrating over all of velocity space, we can also simply substitute $v_{\parallel} \to - v_{\parallel}$ everywhere and the value of the integrals will not change. 

We demonstrate this explicitly with two examples. For $\hat{L}_{\phi \phi}$, we have
\begin{equation}
\begin{split}
    \hat{L}_{\phi \phi}^T  &= \left(\sum_s \frac{Z_s^2 n_{0 s} T_{r}}{n_{r} T_{0 s}} - \int \dd[3]{v} \frac{Z_s^2 F_{0 s} T_r}{n_r T_{0 s}}  J_{0 s} \frac{\omega - n \omega_{\ast s}}{\omega - v_{\parallel} k_{\parallel} - n \omega_{d s}} J_{0 s}\right)^T\\
    & =  \sum_s \frac{Z_s^2 n_{0 s} T_{r}}{n_{r} T_{0 s}} - \int \dd[3]{v} \frac{Z_s^2 F_{0 s} T_r}{n_r T_{0 s}}  \left(J_{0 s} \frac{\omega - n \omega_{\ast s}}{\omega - v_{\parallel} k_{\parallel} - n \omega_{d s}} J_{0 s}\right)^T \\
    & =  \sum_s \frac{Z_s^2 n_{0 s} T_{r}}{n_{r} T_{0 s}} - \int \dd[3]{v} \frac{Z_s^2 F_{0 s} T_r}{n_r T_{0 s}}  J_{0 s} \frac{\omega - n \omega_{\ast s}}{\omega + v_{\parallel} k_{\parallel} - n \omega_{d s}} J_{0 s} \\
    & = \sum_s \frac{Z_s^2 n_{0 s} T_{r}}{n_{r} T_{0 s}} - \int \dd[3]{v} \frac{Z_s^2 F_{0 s} T_r}{n_r T_{0 s}}  J_{0 s} \frac{\omega - n \omega_{\ast s}}{\omega - v_{\parallel} k_{\parallel} - n \omega_{d s}} J_{0 s}\\
    & = \hat{L}_{\phi \phi}. 
\end{split}
\end{equation}
For $\hat{L}_{\phi A}$, we have
\begin{equation}
\begin{split}
    \hat{L}_{\phi A}^T & = \left(\sum_s \int \dd[3]{v} \frac{Z_s^2 F_{0 s} T_r}{n_r T_{0 s} }  J_{0 s} \frac{\omega - n \omega_{\ast s}}{\omega - v_{\parallel} k_{\parallel} - n \omega_{d s}} J_{0 s} \frac{v_{\parallel}}{c_r}\right)^T  \\ 
    & = \sum_s \int \dd[3]{v} \frac{Z_s^2 F_{0 s} T_r}{n_r T_{0 s} }  \left(J_{0 s} \frac{\omega - n \omega_{\ast s}}{\omega - v_{\parallel} k_{\parallel} - n \omega_{d s}} J_{0 s} \frac{v_{\parallel}}{c_r}\right)^T \\ 
    & = \sum_s \int \dd[3]{v} \frac{Z_s^2 F_{0 s} T_r}{n_r T_{0 s} } \frac{v_{\parallel}}{c_r} J_{0 s} \frac{\omega - n \omega_{\ast s}}{\omega + v_{\parallel} k_{\parallel} - n \omega_{d s}} J_{0 s} \\
    & = - \sum_s \int \dd[3]{v} \frac{Z_s^2 F_{0 s} T_r}{n_r T_{0 s} } \frac{v_{\parallel}}{c_r} J_{0 s} \frac{\omega - n \omega_{\ast s}}{\omega - v_{\parallel} k_{\parallel} - n \omega_{d s}} J_{0 s} \\ 
    & = \hat{L}_{A \phi}. 
\end{split}
\end{equation}
It should now be clear by inspection that $\hat{L}$ is indeed complex symmetric by simply reversing the order of all functions and operators of $l$ and taking $v_{\parallel} \to - v_{\parallel}$. 

We now state the variational principle. Let $\bm{\chi}$ be parameterized by complex numbers $\alpha$. We define the functional $p(\omega, \alpha)$ as
\begin{equation}
    p(\omega, \alpha) = \left<\bar{\bm{\chi}}, \hat{L}(\omega) \bm{\chi}\right> = \int_{-\infty}^{\infty} \frac{\dd{l}}{B} \bm{\chi}^T \hat{L}(\omega) \bm{\chi}. 
\end{equation}
We then seek to solve the following equations:
\begin{align}
    p &= 0, \\
    \frac{\partial p}{\partial \alpha} &= 0, \\
    \frac{\partial p}{\partial \omega} &\neq 0. 
\end{align}
Doing so will give us the right eigenfunctions $\bm{\chi}$, the left eigenfunctions $\bar{\bm{\chi}}$, and the eigenvalue $\omega$. 

Note that the only assumptions we have made are 1) the background is Maxwellian, 2) there is no rotation or background electric field $\vb{E}$, 3) there are no collisions, and 4) the ballooning representation correctly describes the local instability. Impurities, arbitrary geometry, and electromagnetic effects are included in the above general formulation. Including a non-trivial collision operator (e.g. pitch-angle scattering) would require expanding the distribution function in terms of basis functions and coupling them together \citep{hamed2018}. On the other hand, the inclusion of $E \times B$ shear through the frequency $\gamma_E$ is problematic. The correct analysis of the problem requires time-dependent $k_\perp$ or magnetic field strengths (depending on the specific representation used), so it is no longer a simple eigenvalue problem \citep{maeyama2025}. Even in the small $\gamma_E$ limit where one just wants an ``instantaneous'' growth rate, the formulation would still break down because the extra differential operators proportional to $\gamma_E$ break the complex symmetry of the gyrokinetic field equations.

Note that we have not explicitly written down the gyrokinetic field equations, but rather neatly represented them to prove complex symmetry. Actually writing down the solutions requires some background knowledge in guiding center dynamics and a detailed analysis of bounce-transit orbits.

\section{A Weak Form of the Gyrokinetic Field Equations}
\label{sec:solutions}
\subsection{Preliminaries} 
All guiding center orbit analysis necessary to write down explicit solutions to the gyrokinetic equation is presented in Appendix~\ref{sec:guiding}. The methods used here are inspired by \citet{tang1980}, \citet{rewoldt1982}, \citet{garbet1990}, and \citet{chandran2024}. The background Maxwellian is
\begin{equation}
    F_{0 s} = \frac{n_{0 s}(\psi)}{\pi^{3/2} v_{T_{s}}^3} \exp(- \varepsilon / T_s(\psi)), 
\end{equation}
where $v_{T_{s}} = \sqrt{2 T_s / m_s}$. We also work in non-orthogonal Clebsch coordinates where the magnetic field $\vb{B}$ can be written as
\begin{equation}
    \vb{B} = \nabla \alpha \times \nabla \psi,
\end{equation}
where $\alpha$ is a field line label and $\psi$ is the poloidal flux normalized by $2 \pi$. We also use a field line following coordinate $\theta$ such that all physical quantities obey $f(\psi, \theta + 2 \pi, \alpha) = f(\psi, \theta, \alpha)$ and $\vb{B} \cdot \nabla \theta = J_{\psi}^{-1} > 0$, where $J_{\psi}$ is the Jacobian of the coordinate system. All perturbed fields are subject to the ballooning transformation:
\begin{equation}
    u(\psi, \theta, \varphi) = \sum_{p} \hat{u}(\psi, \theta + 2 \pi p) e^{i n S(\psi, \theta + 2 \pi p, \varphi)},
\end{equation}
where $\hat{u}$ is the slowly varying envelope, $S$ is the eikonal such that $\vb{B} \cdot \nabla S = 0$, and $n$ is the toroidal mode number.  This guarantees that $u$ is periodic in $\theta$. We will work exclusively in the ballooning representation, so we shall drop the hat for convenience. By convention, $S$ is defined such that
\begin{equation}
    \nabla S = \nabla \alpha + \theta_0 q' \nabla \psi, 
\end{equation}
where $\theta_0$ is the ballooning angle. We define the perpendicular wavevector as
\begin{equation}
    \vb{k}_{\perp} = n \nabla S. 
\end{equation}
Using the coordinates $(\theta, \varepsilon, \mu)$, the linearized gyrokinetic equation can then be written as
\begin{equation}
    \frac{v_{\parallel}}{J_{\psi} B} \frac{\partial h_s}{\partial \theta} - i \left(\omega - n \omega_{d s} \right) h_s = - i \frac{Z_s e}{T_s} F_{0 s} \left(\omega - \omega_{\ast s} \right) J_0 (k_{\perp} \rho_s) \left(\phi - v_{\parallel} A_{\parallel} \right),
\end{equation}
where
\begin{align}
    J_{\psi} &= \left(\nabla \theta \cdot (\nabla \alpha \times \nabla \psi) \right)^{-1}, \\
    n \omega_{d s} & = \vb{k}_{\perp} \cdot \vb{v}_{d s}, \\
    n \omega_{\ast s} &= \frac{n T_{s}}{Z_{s} E} \frac{1}{n_{0 s}} \frac{\partial n_{0 s}}{\partial \psi} \left(1 + \eta_s \left(\frac{\varepsilon}{T_s} - \frac{3}{2} \right)\right), \\
    \rho_s & = \frac{m_s v_{\perp}}{e Z_s B} = \frac{m_s}{e Z_s B} \sqrt{\frac{2 \varepsilon}{m_s}} \sqrt{\lambda B}, 
\end{align}
where the guiding center velocity $\vb{v}_{d}$ is given by
\begin{equation}
    \vb{v}_d = \frac{m}{e B} \frac{v_{\parallel}^2 + \mu B / m}{B^2} \vb{B} \times \nabla B + \frac{m}{e B} \frac{\mu_0 v_{\parallel}^2}{B^3} \vb{B} \times \nabla p,
\end{equation}
where $p$ is the total plasma pressure. For convenience, we have dropped $\delta$ from any perturbed fields. 

\subsection{Passing Particles}
\subsubsection{Passing Solution of the Gyrokinetic Equation}
For the passing solution, we use the boundary condition that $h_s \to 0$ as $|\theta| \to \infty$. We write the solution assuming $\Im \omega > 0$ (otherwise we need to analytically continue the solution). We also need to consider both signs of $v_{\parallel}$ separately. Defining $\theta_{-} = \theta - \theta'$, the solution is
\begin{equation}
    \label{eq:passing_solution}
    h_{s, p, \pm} = \mp i \frac{Z_s e F_{0 s} }{T_s} \left(\omega - n \omega_{\ast s} \right) \int_{\mp \infty}^{\theta} \dd{\theta'} J_{\psi} B J_{0 s} \left(\frac{\phi}{|v_{\parallel}|} \mp A_{\parallel} \right) e^{\pm i I^{\theta}_{\theta'}},
\end{equation}
where $I$ will be defined shortly. Then,
\begin{equation}
    h_{s, p, +} + h_{s, p, -} = -i \frac{Z_s e F_{0 s}}{T_s} \left(\omega - n \omega_{\ast s} \right) \int_{-\infty}^{\infty} \dd{\theta'} J_{\psi} B J_{0 s} \left(\frac{\phi}{|v_{\parallel}|} - \sign(\theta_{-}) A_{\parallel} \right) e^{i \sign(\theta_{-}) I^{\theta}_{\theta'}}. 
\end{equation}
Meanwhile, 
\begin{equation}
    h_{s, p, +} - h_{s, p, -} = -i \frac{Z_s e F_{0 s}}{T_s} \left(\omega - n \omega_{\ast s} \right) \int_{-\infty}^{\infty} \dd{\theta'} J_{\psi} B J_{0 s} \left(\sign(\theta_{-}) \frac{\phi}{|v_{\parallel}|} - A_{\parallel} \right) e^{i \sign(\theta_{-}) I^{\theta}_{\theta'}}. 
\end{equation}
The quantity $I$ is an integral defined as
\begin{equation}
    I_{a}^{b} = \int_{a}^{b} \dd{\theta} \frac{J_{\psi} B}{|v_{\parallel}|} \left(\omega - n \omega_{d s}\right). 
\end{equation}

\subsubsection{Quasineutrality}
For quasineutrality, we will need to integrate
\begin{equation}
    \mathcal{L}_{\phi, s, p} = Z_s e \int_{-\infty}^{\infty} \dd{\theta} J_{\psi} \int \dd[3]{v} \phi J_{0 s} \left(h_{s, p, +} + h_{s, p, -}\right),
\end{equation}
where
\begin{equation}
    \dd[3]{v} = 2 \pi \dd{\varepsilon} \dd{\lambda} \frac{B}{m_s^2} \frac{\varepsilon}{|v_{\parallel}|}
\end{equation}
and the integration limits for passing particles are
\begin{align}
    0 &\le \lambda \le \frac{1}{B_{\text{max}}}, \\
    0 &\le \varepsilon \le \infty. 
\end{align}
It is useful to perform a change of variables using the action angle $\alpha_t$:
\begin{equation}
    \frac{\dd{\theta} J_{\psi} B}{|v_{\parallel}|} = \frac{\dd{\alpha_t}}{\Omega_t}, 
\end{equation}
where $\Omega_t$ is the transit frequency. We also note the following:
\begin{equation}
    I_{0}^{\theta} = \left(\frac{\omega - n \Omega_{d}}{\Omega_t }\right) \alpha_t - n \Lambda_{d s}(\alpha_t), 
\end{equation}
where $\Omega_{d}$ is the transit averaged drift frequency and $\Lambda_{d}$ contains the drift deviations, defined in Eq.~\ref{eq:drift_passing} and Eq.~\ref{eq:deviations_passing} respectively. We then define the following Fourier transform pairs:
\begin{align}
    \phi^{+}_{s}(k) &= \int_{-\infty}^{\infty} \frac{\dd{\alpha_t}}{2 \pi}  J_{0 s} \phi e^{-i k \alpha_t + i n \Lambda_{d s}(\alpha_t)}, \\
    \phi^{-}_{s}(k) &= \int_{-\infty}^{\infty} \frac{\dd{\alpha_t}}{2 \pi}  J_{0 s} \phi e^{i k \alpha_t - i n \Lambda_{d s}(\alpha_t)}, \\
    A^{+}_{s}(k) &= \int_{-\infty}^{\infty} \frac{\dd{\alpha_t}}{2 \pi}  J_{0 s} |v_{\parallel}| A_{\parallel} e^{-i k \alpha_t + i n \Lambda_{d s}(\alpha_t)}, \\
    A^{-}_{s}(k) &= -\int_{-\infty}^{\infty} \frac{\dd{\alpha_t}}{2 \pi}  J_{0 s} |v_{\parallel}| A_{\parallel} e^{i k \alpha_t - i n \Lambda_{d s}(\alpha_t)}, \\
    J_{0 s} \phi e^{i n \Lambda_{d s}(\alpha_t)} & = \int_{-\infty}^{\infty} \dd{k} e^{i k \alpha_t} \phi^{+}_{s}(k), \\
    J_{0 s} \phi e^{-i n \Lambda_{d s}(\alpha_t)} & = \int_{-\infty}^{\infty} \dd{k} e^{-i k \alpha_t} \phi^{-}_{s}(k), \\
    J_{0 s} |v_{\parallel}| A_{\parallel} e^{i n \Lambda_{d s}(\alpha_t)} & = \int_{-\infty}^{\infty} \dd{k} e^{i k \alpha_t} A^{+}_{s}(k), \\ 
    J_{0 s} |v_{\parallel}| A_{\parallel} e^{-i n \Lambda_{d s}(\alpha_t)} & = -\int_{-\infty}^{\infty} \dd{k} e^{-i k \alpha_t} A^{+}_{s}(k).
\end{align}
We then obtain the following identities:
\begin{equation}
\begin{split}
    \int_{-\infty}^{\infty} \dd{\theta} \frac{J_{\psi} B}{|v_{\parallel}|} \int_{-\infty}^{\infty} \dd{\theta'} \frac{J_{\psi}' B'}{|v_{\parallel}|'} J_{0 s} \phi J_{0 s}' \phi' e^{i \sign(\theta_{-}) I_{\theta'}^{\theta}} \\ =  \frac{4 \pi i}{\Omega_{t s}} \int_{-\infty}^{\infty} \dd{k} \frac{\phi^{+}_{s}(k) \phi^{-}_{s}(k)}{\omega - k \Omega_{t s} - n \Omega_{d s}} 
\end{split}
\end{equation}
and
\begin{equation}
\begin{split}
    \int_{-\infty}^{\infty} \dd{\theta} \frac{J_{\psi} B}{|v_{\parallel}|} \int_{-\infty}^{\infty} \dd{\theta'} J_{\psi}'B' \sign(\theta_{-}) J_{0 s} \phi J_{0 s}' A_{\parallel}'  e^{i \sign(\theta_{-}) I_{\theta'}^{\theta}} \\ = \frac{2 \pi i}{\Omega_{t s}} \int_{-\infty}^{\infty} \dd{k} \frac{A^{+}_{s}(k) \phi^{-}_{s}(k) + A^{-}_{s}(k) \phi^{+}_{s}(k)}{\omega - k \Omega_{t s} - n \Omega_{d s}}.
\end{split}
\end{equation}

Substituting this into the passing solution, we find
\begin{equation}
     \mathcal{L}_{\phi, p, s} = \frac{2 \pi^2 Z_s^2 e^2}{m_s^2 T_s} \int_{0}^{\infty} \dd{\varepsilon} \varepsilon F_{0 s} (\omega - 
     n \omega_{\ast s}) \int_0^{1 / B_{\text{max}}} \frac{\dd{\lambda}}{\Omega_{t s}} \int_{-\infty}^{\infty} \dd{k} \frac{2 \phi^{+}_{s} \phi^{-}_{s} - A^{+}_{s} \phi^{-}_{s} - A^{-}_{s} \phi^{+}_{s}}{\omega - k \Omega_{t s} - n \Omega_{d s}}. 
\end{equation}
This compactly takes into account both signs of the parallel velocity. As noted in \citet{garbet1990}, this form is more numerically advantageous than the original form. In the low $\varepsilon$ limit, the exponentials in Eq.~\ref{eq:passing_solution} are highly oscillatory due to the term proportional to $\omega / |v_{\parallel}|$. We have moved this term into the Landau resonance. Meanwhile, the remaining eikonals in the Fourier transform pairs are well behaved as $\varepsilon \to 0$, since $\omega_{d} / |v_{\parallel}| \propto \sqrt{\varepsilon}.$

\subsubsection{Ampere's Law}
For Ampere's law, we will need to integrate
\begin{equation}
    \mathcal{L}_{A, p, s} = Z_s e \int_{-\infty}^{\infty} \dd{\theta} \int \dd[3]{v} A_{\parallel} |v_{\parallel}| J_{0 s} \left(h_{s, p, +} - h_{s, p, -}\right). 
\end{equation}
The steps are similar to that of quasineutrality, so we just write the final result:
\begin{equation}
     \mathcal{L}_{A, p, s} = \frac{2 \pi^2 Z_s^2 e^2}{m_s^2 T_s} \int_{0}^{\infty} \dd{\varepsilon} \varepsilon F_{0 s} (\omega - n \omega_{\ast s}) \int_0^{1 / B_{\text{max}}} \frac{\dd{\lambda}}{\Omega_{t s}} \int_{-\infty}^{\infty} \dd{k} \frac{2 A^{+}_{s} A^{-}_{s} - A^{+}_{s} \phi^{-}_{s} - A^{-}_{s} \phi^{+}_{s}}{\omega - k \Omega_{t s} - n \Omega_{d s}}. 
\end{equation}

\subsection{Trapped Particles}
\subsubsection{Trapped Solution of the Gyrokinetic Equation}
For the trapped solution, we instead need to use the boundary conditions 
\begin{align}
    h_{s, tr, +} (\theta_1) &= h_{s, tr, -} (\theta_1), \\
    h_{s, tr, +} (\theta_2) &= h_{s, tr, -} (\theta_2),
\end{align}
where $\theta_1$ and $\theta_2$ are bounce points such that $\theta_1 < \theta < \theta_2$. This leads to
\begin{equation}
\label{eq:trapped_solution_1}
\begin{split}
    h_{s, tr, +} + h_{s, tr, -} = \\ \frac{2 Z_s e F_{0 s}}{T_s} \frac{\omega - 
    n \omega_{\ast s}}{\sin (I_{\theta_1}^{\theta_2}) } \Bigg[ & \int_{\theta_1}^{\theta} \dd{\theta'} J_{\psi} B J_{0 s} \left(\frac{\phi}{|v_{\parallel}|} \cos(I_{\theta_1}^{\theta'}) \cos(I_{\theta}^{\theta_2}) + i A_{\parallel} \sin(I_{\theta_1}^{\theta'}) \cos(I_{\theta}^{\theta_2}) \right)  \\ & + \int_{\theta}^{\theta_2} \dd{\theta'} J_{\psi} B J_{0 s} \left(\frac{\phi}{|v_{\parallel}|} \cos(I_{\theta_1}^{\theta}) \cos(I_{\theta'}^{\theta_2}) - i A_{\parallel} \cos(I_{\theta_1}^{\theta}) \sin(I_{\theta'}^{\theta_2})  \right)\Bigg],
\end{split}
\end{equation}
and
\begin{equation}
\label{eq:trapped_solution_2}
\begin{split}
    h_{s, tr, +} - h_{s, tr, -} = \\ \frac{2 Z_s e F_{0 s}}{T_s} \frac{\omega - 
    n \omega_{\ast s}}{\sin (I_{\theta_1}^{\theta_2}) } \Bigg[ & \int_{\theta_1}^{\theta} \dd{\theta'} J_{\psi} B J_{0 s} \left(- \frac{\phi}{|v_{\parallel}|}\cos(I_{\theta_1}^{\theta'}) \sin(I_{\theta}^{\theta_2}) + i A_{\parallel} \sin(I_{\theta_1}^{\theta'}) \sin(I_{\theta}^{\theta_2}) \right)  \\ & + \int_{\theta}^{\theta_2} \dd{\theta'} J_{\psi} B J_{0 s} \left(\frac{\phi}{|v_{\parallel}|} \sin(I_{\theta_1}^{\theta}) \cos(I_{\theta'}^{\theta_2}) + i A_{\parallel} \sin(I_{\theta'}^{\theta_2}) \sin(I_{\theta}^{\theta_1}) \right)\Bigg]. 
\end{split}
\end{equation}
Note that we need to pick $\theta_1, \theta_2$ such that $\theta_1 \le \theta \le \theta_2$.

\subsubsection{Quasineutrality}
Integrating over the trapped part of velocity space requires much more work than for passing particles. As before, we have
\begin{equation}
    \dd[3]{v} = 2 \pi \dd{\varepsilon} \dd{\lambda} \frac{B}{m_s^2} \frac{\varepsilon}{|v_{\parallel}|}. 
\end{equation}
The limits of integration for trapped particles, however, are
\begin{align}
    \frac{1}{B_{\text{max}}} &\le \lambda \le \frac{1}{B(\theta)}, \\
    0 & \le \varepsilon \le \infty. 
\end{align}
Thus, the upper limit of the pitch angle parameter depends on the current position along the field line. We can, however, use a trick to make the integration limits more manageable. Let $\theta_{\text{min}} \le \theta_1 \le \theta_2 \le \theta_{\text{max}}$, where $\theta_{\text{min}}$ and $\theta_{\text{max}}$ correspond to the extent of the bounce well. Then,
\begin{equation}
    \int_{\theta_{\text{min}}}^{\theta_{\text{max}}} \dd{\theta} \int_{1/ B_{\text{max}}}^{1/ B(\theta)} \dd{\lambda} = \int_{1/B_{\text{max}}}^{1/B_{\text{min}}} \dd{\lambda} \int_{\theta_1}^{\theta_2} \dd{\theta}. 
\end{equation}
We also note the following identity:
\begin{equation}
\begin{split}
    \int_{\theta_1}^{\theta_2} \dd{\theta} \frac{J_{\psi} B}{|v_{\parallel}|} & J_{0 s} \phi \int_{\theta_1}^{\theta} \dd{\theta'} \frac{J_{\psi}' B'}{|v_{\parallel}|'} J_{0 s}' \phi'\cos(I_{\theta_1}^{\theta'}) \cos(I_{\theta}^{\theta_2})  \\ &+ \int_{\theta_1}^{\theta_2} \dd{\theta} \frac{J_{\psi} B}{|v_{\parallel}|} J_{0 s} \phi \int_{\theta}^{\theta_2} \dd{\theta'} \frac{J_{\psi}' B'}{|v_{\parallel}|'} J_{0 s}' \phi' \cos(I_{\theta_1}^{\theta}) \cos(I_{\theta'}^{\theta_2}) \\ & = \frac{\pi  \sin(I_{\theta_1}^{\theta_2}) }{\Omega_{b s}} \sum_{n_2 = -\infty}^{\infty} \frac{\left(\phi^{n_2}_{s}\right)^2}{\omega - n_2 \Omega_{b s}- n \Omega_{d s}}, 
\end{split}
\end{equation}
where $\phi^{n_2}_{s}$ is defined as
\begin{equation}
    \phi^{n_2}_{s} = \left< J_{0 s} \phi \cos(n \tilde{w}_{d s} - n_2 \alpha_b)\right>_b. 
\end{equation}
Here, $\Omega_{d}$ and $\tilde{w}_{d}$ are the bounce averaged drift frequency and drift deviations, defined in Eqs.~\ref{eq:drift_trapped}-\ref{eq:deviations_trapped}. We are essentially computing the bounce harmonics of the electrostatic potential modulated by finite Larmor radius (FLR) effects via the Bessel function and also finite banana width effects via $\tilde{w}_{d s}$. Dropping the $s$ subscript momentarily, one can prove this by writing
\begin{align}
    J_{0} \phi \cos(n \tilde{w}_{d}) &= \phi_0^c + \sum_{n_2=1}^{\infty} 2 \phi_{n_2}^{c}\cos(n_2 \alpha_b) , \\
    J_{0} \phi \sin(n \tilde{w}_{d}) &= \sum_{n_2=1}^{\infty} 2 \phi_{n_2}^{s} \sin(n_2 \alpha_b),
\end{align}
as well as
\begin{align}
    \phi^{n_2} &= \phi_{n_2}^{c} + \phi_{n_2}^{s}, \\
    \phi^{-n_2} &= \phi_{n_2}^{c} - \phi_{n_2}^{s},
\end{align}
and computing the integrals manually. Meanwhile, we define a similar bounce average for the electromagnetic piece:
\begin{equation}
    A^{n_2}_{s} = \left<i J_{0 s} v_{\parallel} A_{\parallel} \sin(n \tilde{w}_{d s}  - n_2 \alpha_b )\right>_b.
\end{equation}
(Note the factor of $i$.) One can then write (again dropping the $s$ subscript for clarity)
\begin{align}
    i J_{0} |v_{\parallel}| A_{\parallel} \sin(n \tilde{w}_{d}) & = A_0^c + \sum_{n_2=1}^{\infty} 2 A_{n_2}^c \cos(n_2 \alpha_b), \\
    i J_{0} |v_{\parallel}| A_{\parallel} \cos(n \tilde{w}_{d} ) & = \sum_{n_2 = 1}^{\infty} 2 A_{n_2}^s \sin(n_2 \alpha_b),
\end{align}
as well as
\begin{align}
    A^{n_2} &= A_{n_2}^c - A_{n_2}^s, \\
    A^{-n_2} &= A_{n_2}^c + A_{n_2}^s. 
\end{align}
This leads to the following identity:
\begin{equation}
\begin{split}
    \int_{\theta_1}^{\theta_2} \dd{\theta} \frac{J_{\psi} B}{|v_{\parallel}|} & J_{0 s} \phi \int_{\theta_1}^{\theta} \dd{\theta'} J_{\psi}' B' J_{0 s}' i A_{\parallel}' \sin(I_{\theta_1}^{\theta'}) \cos(I_{\theta}^{\theta_2}) 
    \\ & - \int_{\theta_1}^{\theta_2} \dd{\theta} \frac{J_{\psi} B}{|v_{\parallel}|} J_{0 s} \phi \int_{\theta}^{\theta_2} \dd{\theta'} J_{\psi}' B' J_{0 s}' i A_{\parallel}'  \cos(I_{\theta_1}^{\theta}) \sin(I_{\theta'}^{\theta_2}) \\  &=- \frac{\pi \sin(I_{\theta_1}^{\theta_2}) }{ \Omega_{b s}}  \sum_{n_2 = -\infty}^{\infty} \frac{\phi^{n_2}_{s} A^{n_2}_{s}}{\omega - n_2 \Omega_{b s}- n \Omega_{d s}} 
\end{split}
\end{equation}

We can finally write down the trapped part of the integral we need to compute. First, define
\begin{equation}
    \mathcal{L}_{\phi, tr, s} = Z_s e \int_{-\infty}^{\infty} \dd{\theta} J_{\psi} \int \dd[3]{v} \phi J_{0 s} \left(h_{s, tr, +} + h_{s, p, -}\right).
\end{equation}
Then, we find
\begin{equation}
    \mathcal{L}_{\phi, tr, s} = \sum_p \sum_{n_2} \frac{4 \pi^2 Z_s^2 e^2}{m_s^2 T_s} \int_{0}^{\infty} \dd{\varepsilon} \varepsilon F_{0 s} (\omega - 
    n \omega_{\ast s} )\int_{1/B_{\text{max}}}^{1/B_{\text{min}}} \frac{\dd{\lambda}}{\Omega_{b s}} \frac{\phi^{n_2}_{s} (\phi^{n_2}_{s} - A^{n_2}_{s})}{\omega - n_2 \Omega_{b s} - n \Omega_{d s}},
\end{equation}
where the sum over $p$ denotes a sum over all bounce wells along the field line. Note that we are also summing over all bounce harmonics via $n_2$. As in the passing case, the exponentials Eq.~\ref{eq:trapped_solution_1}-\ref{eq:trapped_solution_2} are highly oscillatory due to the term proportional to $\omega / |v_{\parallel}|$. We have moved this term into the Landau resonance. Meanwhile, the remaining eikonals in the Fourier transform pairs are well behaved as $\varepsilon \to 0$, since $\omega_{d} / |v_{\parallel}| \propto \sqrt{\varepsilon}.$

\subsubsection{Ampere's Law}
The process for calculating
\begin{equation}
    \mathcal{L}_{A, tr, s} = Z_s e \int_{-\infty}^{\infty} \int \dd[3]{v} |v_{\parallel}| A_{\parallel} J_{0 s} \left(h_{s, tr, +} - h_{s, tr, -}\right)
\end{equation}
is similar, so we quote the final result:
\begin{equation}
    \mathcal{L}_{A, tr, s} = \sum_p \sum_{n_2} \frac{4 \pi^2 Z_s^2 e^2}{m_s^2 T_s} \int_{0}^{\infty} \dd{\varepsilon} \varepsilon F_{0 s} (\omega - n \omega_{\ast s} )\int_{1/B_{\text{max}}}^{1/B_{\text{min}}} \frac{\dd{\lambda}}{\Omega_{b s}} \frac{A^{n_2}_{s} (A^{n_2}_{s} - \phi^{n_2}_{s})}{\omega - n_2 \Omega_{b s} - n \Omega_{d s}}
\end{equation}

\subsection{Full Solution and Weak Form}
The quasineutrality equation and parallel Ampere's law read:
\begin{align}
    \sum_s \frac{n_{0 s}^2 Z_s^2 e^2 }{T_s} \phi - \sum_s Z_s e \int \dd[3]{v} J_{0 s} (h_{s, +} + h_{s, -} )&= 0, \\
    \frac{k_{\perp}^2}{\mu_0} A_{\parallel} - \sum_s Z_s e \int \dd[3]{v} J_{0 s} |v_{\parallel}| (h_{s, +} - h_{s, -} ) & = 0. 
\end{align}
The proper variational principle is to define the integral
\begin{equation}
\begin{split}
    \mathcal{L} = \int_{-\infty}^{\infty} \dd{\theta} J_{\psi} \phi \left(\sum_s \frac{n_{0 s}^2 Z_s^2 e^2 }{T_s} \phi + \sum_s Z_s e \int \dd[3]{v} J_{0 s} (h_{s, +} + h_{s, -} ) \right)  \\ +  \int_{-\infty}^{\infty} \dd{\theta} J_{\psi} A_{\parallel} \left( \frac{k_{\perp}^2}{\mu_0} A_{\parallel} - \sum_s Z_s e \int \dd[3]{v} J_{0 s} |v_{\parallel}| (h_{s, +} - h_{s, -} )\right). 
\end{split}
\end{equation}
We have already computed the resonant components. We define the non-resonant pieces as
\begin{equation}
    \mathcal{L}_{0} = \sum_s \frac{n_{0 s}^2 Z_s^2 e^2 }{T_s} \int_{-\infty}^{\infty} \dd{\theta} J_{\psi} \phi^2 + 
    \int_{-\infty}^{\infty} \dd{\theta} J_{\psi} \frac{k_{\perp}^2}{\mu_0} A_{\parallel}^2. 
\end{equation}
Then we simply write 
\begin{equation}
    \mathcal{L} = \mathcal{L}_{0} - \sum_s \mathcal{L}_{p, s} - \mathcal{L}_{tr, s}, 
\end{equation}
where
\begin{equation}
\begin{split}
    \mathcal{L}_{p, s} & = \mathcal{L}_{\phi, p, s} + \mathcal{L}_{A, p, s} \\ & = \frac{4\pi^2 Z_s^2 e^2}{m_s^2 T_s} \int_{0}^{\infty} \dd{\varepsilon} \varepsilon F_{0 s} \int_0^{1 / B_{\text{max}}} \frac{\dd{\lambda}}{\Omega_{t s}} \int_{-\infty}^{\infty} \dd{k} \frac{\left(\omega - n \omega_{\ast s} \right) \left(\phi^{+}_{s} - A^{+}_{s} \right) \left(\phi^{-}_{s} - A^{-}_{s} \right) }{\omega - k \Omega_{t s} - n \Omega_{d s}}
\end{split}
\end{equation}
and
\begin{equation}
\begin{split}
    \mathcal{L}_{tr, s} & = \mathcal{L}_{\phi, tr, s} + \mathcal{L}_{A, tr, s} \\ & = \sum_{p} \sum_{n_2} \frac{4 \pi^2 Z_s^2 e^2}{m_s^2 T_s} \int_{0}^{\infty} \dd{\varepsilon} \varepsilon F_{0 s} \int_{1/B_{\text{max}}}^{1/B_{\text{min}}} \frac{\dd{\lambda}}{\Omega_{b s}} \frac{\left(\omega - n \omega_{\ast s} \right)\left(\phi^{n_2}_{s} - A^{n_2}_{s}\right)^2 }{\omega - n_2 \Omega_{b s} - n \Omega_{d s}}. 
\end{split}
\end{equation}
In effect, we are just computing transit and bounce averages of the potentials along the field line, taking into account FLR effects and drift deviations. The form for passing particles and trapped particles are essentially the same, except that for trapped particles we discretely sum over all bounce harmonics and bounce wells. In contrast, for passing particles we must compute a continuum integral in Fourier space. 

If we parameterize $(\phi, A_{\parallel})$ with tuples of complex numbers $\alpha$, then the variational principle is
\begin{align}
    \mathcal{L}(\omega, \alpha) &= 0, \\
    \frac{\partial \mathcal{L}}{\partial a} (\omega, \alpha) & = 0. 
\end{align}
If we assume local analyticity of $\omega$ in terms of $\alpha$, this essentially implies the following stationary principle:
\begin{equation}
    \frac{\partial \omega}{\partial \alpha} = 0. 
\end{equation}
One can see this and naively assume a min-max principle applies as in quantum mechanics. For example, one could think that the growth rate $\gamma$ might be maximized at the solution. However, we note that by the maximum modulus principle, the appropriate solution will generally correspond to $\gamma(\alpha)$ being a saddle point. Essentially, because the eigenvalues are complex and we have a non-Hermitian system, we cannot guarantee a min-max principle for the growth rate. 

Finally, we point out that we can extend the analysis to include $\delta B_{\parallel}$ by considering the parallel part of Ampere's law and including the appropriate terms as seen in Sec.~\ref{sec:variational_intro}.

\section{Conclusions}
\label{sec:conclusions}
In this work, we have corrected the variational principle for local linear collisionless gyrokinetics. In doing so, we have also derived a rigorous weak form for the field equations in general geometry and with electromagnetic effects with no other approximations. While the methods in \citep{rewoldt1982} and \citep{garbet1990} were originally designed for tokamak geometry, we have extended them to account for (periodized) stellarator geometry. Moreover, we have modified the methods and integral transforms in the preceding works to produce more convenient expressions for passing particles and also allow for the variation of nonlinear coefficients for any set of suitable trial functions. For example, rather than setting a constant Gaussian width as in \citep{rewoldt1982}, one can now directly vary the width itself per the variational principle (as one often does in introductory quantum mechanics). 

The equations as written can be further reduced and modified with various approximations, as done in QuaLiKiz \citep{stephens2021}. Moreover, there exist two obvious theoretical extensions. The first would be to include a non-trivial collision operator, such as pitch angle scattering. Such a method would require that the distribution function be expanded in terms of basis functions, such as eigenfunctions of the collision operator. One would then obtain coupled linear differential equations for components of the distribution function. Then, one would either need to prove complex symmetry of the resulting field equations or else derive another variational principle. The second extensions would be to relax the local assumption and examine global eigenmodes with a finite radial width. It would be interesting to determine if a similar variational principle exists for global eigenmodes. 

The derived equations can also be used to develop codes to solve the eigenvalue problem with fewer assumptions. For example, QuaLiKiz assumes a shifted circular geometry and does not include electromagnetic effects. A variational code that included both electromagnetic effects and general geometry would prove advantageous in simulating microinstabilities in stellarator geometries. Moreover, the variational approach allows for a self-consistent determination of eigenvalues and eigenfunctions without requiring a completely predetermined trial function. 

Lastly, one could also develop codes that utilize state-of-the-art finite element techniques to obtain fast but highly accurate eigenvalues and eigenmodes. A conventional approach to solving the eigenvalue equation (such as in the gyrokinetic code GENE \citep{jenko2000}) is to employ a fixed finite difference discretization in phase space (including velocity space). In contrast, the above variational approach discretizes velocity space via numerical integration, which generically can be adaptive. Moreover, when discretizing the field line following coordinate via a polynomial finite element method, one can adaptively refine the mesh with a combination of h-refinement (subdividing the elements) and p-refinement (using higher-order polynomials), called hp refinement \citep{finite_element_analysis2021}. Moreover, since we only need to consider finite elements along a single dimension (the field line following coordinate), the number of refinement candidates would be relatively modest.

\section*{Acknowledgments}
The authors thank Joshua Burby, Norman Cao, William Barham, and Max Ruth for insightful discussions. We also thank David Hatch and Swadesh Mahajan for providing useful references. 

\section*{Declaration of Interests}
The authors report no conflict of interest.

\section*{Funding}
This work was supported by the U.S. Department of Energy under Grant No.~DE-FG02-04ER54742 (IFS). 

\appendix
\clearpage

\section{Guiding Center Orbit Analysis}
\label{sec:guiding}

\subsection{Preliminaries}

To write down an explicit solution to the gyrokinetic equation in variational form, it is most convenient to take advantage of various integral transforms. These integral transforms require solving the guiding center equations to zeroth and first order. These ordinary differential equations can be solved ahead of time for a given geometry with a numerical integrator, and their solutions define the integral transforms used in Sec.~\ref{sec:solutions}. Using the Clebsch coordinates defined in Sec.~\ref{sec:solutions}, the guiding center equations of motion are then
\begin{align}
    \frac{\dd{\psi}}{\dd{t}} & = \vb{v}_{d} \cdot \nabla \psi, \\
    \frac{\dd{\alpha}}{\dd{t}} & = \vb{v}_{d} \cdot \nabla \alpha, \\
    \frac{\dd{\theta}}{\dd{t}} & = v_{\parallel} \vu{b} \cdot \nabla \theta + \vb{v}_{d} \cdot \nabla \theta, \\
    \frac{\dd{v_{\parallel}}}{\dd{t}} & = - \frac{\mu}{m} \vu{b} \cdot \nabla B,
\end{align}
where $t$ represents time. The magnetic drift velocity $\vb{v}_{d}$ is given by
\begin{equation}
    \frac{m}{e B} \frac{v_{\parallel}^2 + \mu B / m}{B^2} \vb{B} \times \nabla B + \frac{m}{e B} \frac{\mu_0 v_{\parallel}^2}{B^3} \vb{B} \times \nabla p,
\end{equation}

Both the energy $\varepsilon$ and the magnetic moment $\mu$ are constants of motion. We note that $\psi$ and $\alpha$ only change due to the drift motion, whereas $\theta$ changes both due to the parallel motion and the drift motion. The parallel velocity in turn changes due to the mirror force. Also, note that
\begin{equation}
    \vu{b} \cdot \nabla f = \frac{1}{J_{\psi} B} \frac{\partial f}{\partial \theta}. 
\end{equation}
It will be convenient to define the pitch angle parameter,
\begin{equation}
    \lambda = \frac{\mu}{\varepsilon}, 
\end{equation}
which is also a constant of motion. This allows us to write the parallel velocity as
\begin{equation}
    v_{\parallel}^2 = \frac{2 \varepsilon}{m} \left(1 - \lambda B \right). 
\end{equation}

Rather than solve these equations of motion self-consistently, it is advantageous to separate the system into fast timescales and slow timescales. This is much like how the fast cyclotron motion was averaged out to obtain the guiding center equations of motion. Therefore, we solve the equations perturbatively: the fast field-line motion is determined by 
\begin{align}
    \frac{\dd{\psi}}{\dd{t}} & = 0, \\
    \frac{\dd{\alpha}}{\dd{t}} & = 0, \\
    \frac{\dd{\theta}}{\dd{t}} & = v_{\parallel} \vu{b} \cdot \nabla \theta, \\
    \frac{\dd{v_{\parallel}}}{\dd{t}} & = - \frac{\mu}{m} \vu{b} \cdot \nabla B. 
\end{align}
These are the zeroth order guiding center equations of motion. The first two equations can be trivially integrated out as $\psi = \psi_0$ and $\alpha = \alpha_0$, which determine the reference flux surface and the reference field line. To lowest order, the particle is bound to the field line according to the mirror force. All magnetic field quantities are evaluated at $\psi_0, \alpha_0$, meaning that they are purely functions of $\theta.$ This reduced system is integrable and Hamiltonian, where the Hamiltonian $H$ is
\begin{equation}
    H = \frac{p_{\theta}^2}{2 m J_{\psi}^2 B^2} + \mu B. 
\end{equation}
Notice that the kinetic term depends on $theta$. This Hamiltonian can also be put into a separable form (meaning $H = T(p) + V(q)$), which is convenient for certain ODE integrators. Define the arc length $l$ along the field line as
\begin{equation}
    l(\theta) = \int_0^{\theta} \dd{\theta'} J_{\psi} B,
\end{equation}
meaning $\vu{b} \cdot \nabla l = 1$ and $v_{\parallel} = \dot{l}$. The Hamiltonian can then be rewritten as
\begin{equation}
    H' = \frac{p^2}{2 m} + \mu B, 
\end{equation}
where $B$ is now a function of $l$. The Hamiltonian is now separable. 

When analyzing the field-line motion, there are two types of particles: trapped particles and passing (circulating) particles. A trapped particle is trapped in a magnetic well and thus periodically bounces between two mirror points, akin to a pendulum; this is termed a bounce orbit. At these mirror points, $v_{\parallel} = 0$ and changes sign. A passing particle, instead, has just the right pitch angle that $|v_{\parallel}| > 0$ always, meaning it never changes direction; this is termed a transit orbit. It is similar to giving a pendulum enough kinetic energy such that it goes round and round over the top. Therefore, it is convenient to analyze trapped particles and passing particles separately. 

To first order, we calculate the magnetic drift velocity and evaluate it at $(\psi_0, \theta(t), \alpha_0)$. We then solve
\begin{align}
    \frac{\dd{\tilde{\psi}}} {\dd{t}} &= \vb{v}_{d} \cdot \nabla \psi \Big|_{\psi_0, \theta(t), \alpha_0}, \\
    \frac{\dd{\tilde{\alpha}}}{\dd{t}} & = \vb{v}_{d} \cdot \nabla \alpha \Big|_{\psi_0, \theta(t), \alpha_0}. 
\end{align}
Here, $\tilde{\psi}$ and $\tilde{\alpha}$ are the deviations from the flux surface and field line, respectively. These are first order equations of motion where the right-hand side is purely a function of $t$, so they can be directly integrated. For omnigenous fields, the radial deviations will tend to average to zero over short time scales \citep{helander2014}, whereas the field line deviations will not. This means that the particle will slowly precess around the flux surface. Over very long time scales, the particle may end up on a different field line on the flux surface, in turn affecting the bounce-transit dynamics. So, we expect this analysis to only hold on time scales shorter than the drift time. Moreover, it only holds if $\rho^\ast \ll 1$, where $\rho^\ast$ is the gyroradius divided by some macroscopic length scale, such as the machine size; this is because $v_{\parallel}/v_D \sim \rho^\ast$. 

For certain geometries, the solutions to $\tilde{\psi}$ and $\tilde{\alpha}$ will consist of a secular piece and a periodic piece. We will denote the periodic piece of each as $\hat{\psi}$ and $\hat{\alpha}$ respectively when appropriate.

\subsection{Passing Particle Orbits}
\label{sec:passing}

\subsubsection{Zeroth Order Transit Motion}
Passing particles have to be treated slightly differently in stellarators vs. tokamaks; the periodicity of the magnetic field in a tokamak makes the motion along the file line essentially periodic (modulo poloidal turns), which is not the case for a general stellarator geometry. Therefore, certain details apply only to tokamak geometries. 

A passing particle has non-vanishing parallel velocity, meaning that along the field $v_{\parallel}$ never changes sign: 
\begin{equation}
    v_{\parallel} = \pm \sqrt{\frac{2 \varepsilon}{m}} \sqrt{1 - \lambda B}. 
\end{equation}
Let $B_{\text{max}}$ denote the maximum magnetic field strength along the field line. The particle is then passing if
\begin{equation}
    0 \le \lambda \le \frac{1}{B_{\text{max}}}. 
\end{equation}
Thus, $\lambda = 0$ corresponds to a deeply passing particle that does not feel the mirror force whatsoever, and $\lambda = 1/B_{\text{max}}$ corresponds to a barely passing particle that rests in unstable equilibrium at the top of the global magnetic field well. 

Because the magnetic field is periodic in a tokamak, the transit motion is essentially periodic. The transit period is defined as
\begin{equation}
    \tau_t = \oint \dd{\theta} \left(\frac{\dd{\theta}}{\dd{t}}\right)^{-1} = \sqrt{\frac{m}{2 \varepsilon}} \int_{-\pi}^{\pi} \frac{J_{\psi} B \dd{\theta}}{\sqrt{1 - \lambda B}}. 
\end{equation}
Since all quantities in the integrand are $2 \pi$ periodic, the limits of the integral do not matter as long as the upper limit is $2 \pi$ greater than the lower limit. This is analogous to finding the period of a pendulum where the bob can swing through the top. It is convenient to define a dimensionless transit time:
\begin{equation}
    \hat{\tau}_t =  \frac{1}{q R_0} \int_{-\pi}^{\pi} \frac{J_{\psi} B \dd{\theta}}{\sqrt{1 - \lambda B}}. 
\end{equation}
The transit frequencies (both with and without dimensions) are defined as
\begin{align}
    \Omega_t &= \frac{2 \pi}{\tau_t}, \\
    \hat{\Omega}_t &= \frac{2 \pi}{\hat{\tau}_t}. 
\end{align}

Solving for the transit motion as a function of time introduces redundancies. If one varies the energy without varying $\lambda$, then one obtains the same motion, just on a different time scale. (This is similar to adjusting the length of a pendulum and adjusting the kinetic energy to compensate.) We therefore introduce the action angle variable $\alpha_t$, defined as
\begin{equation}
    \frac{\dd{\alpha}_t}{\dd{t}} = \Omega_t = \frac{1}{q R_0} \sqrt{\frac{2 \varepsilon}{m}} \hat{\Omega}_t. 
\end{equation}
We can then rewrite the second-order equation of motion as
\begin{equation}
    \frac{\dd}{\dd{\alpha_t}} \left(\frac{J_\psi B}{q R_0} \frac{\dd{\theta}}{\dd{\alpha_t}}\right) = - \frac{q R_0}{J_\psi B} \frac{\lambda}{2 \hat{\Omega}_t^2} \frac{\partial B}{\partial \theta}. 
\end{equation}
The modified Hamiltonian is then
\begin{equation}
    \hat{H}(\theta, \hat{p}_{\theta}) = \frac{1}{2} \left(\frac{\hat{p}_{\theta} q R_0}{J_{\psi} B}\right)^2 + \frac{\lambda B}{2 \hat{\Omega}_2^2}.  
\end{equation}
Then, $\theta(\alpha_t)$ will be
\begin{equation}
    \theta(\alpha_t) = \alpha_t + \tilde{\theta}(\alpha_t), 
\end{equation}
where $\tilde{\theta}$ is $2 \pi$ periodic. So the solution is the sum of a linear secular term and a periodic term. At this point, we can solve for $\alpha_t$ via any ODE integrator. Note that if we wish to make the Hamiltonian separable, that is of the form
\begin{equation}
    H(l, p_l) = \frac{p_l^2}{2} + \frac{f(l)}{2}. 
\end{equation}
once can transform to the normalized arc length $l$ via 
\begin{equation}
    l(\theta) =\frac{1}{q R_0} \int_0^{\theta} \dd{\theta}' J_{\psi} B, 
\end{equation}
in which case
\begin{equation}
    \hat{H} = \frac{p_l^2}{2} + \frac{\lambda B}{2 \hat{\Omega}_2^2}. 
\end{equation}

The initial conditions turn out to be arbitrary. For simplicity, we set
\begin{align}
    \theta(\alpha_t = 0) &= 0, \\
    \frac{\dd{\theta}}{\dd{\alpha_t}}(\alpha_t = 0) &= \left(\frac{q R_0}{J_{\psi} B}  \frac{\sqrt{1 - \lambda B}}{\hat{\Omega}_t} \right) \Bigg|_{\theta = 0} 
\end{align}
The second initial condition can be obtained by considering we want
\begin{equation}
    \frac{\dd{\theta}}{{\dd{t}}}(t=0) = \left( \sqrt{\frac{2 \varepsilon}{m}} \frac{\sqrt{1 - \lambda B}}{J_{\psi} B} \right) \Bigg|_{\theta = 0}. 
\end{equation}
Essentially, we are setting $v_{\parallel}$ at $\theta = 0$ to be consistent with our choice in $\varepsilon$ and $\lambda$. Due to time reversibility and periodicity in tokamak geometry, we do not need to separately consider passing particles traveling backwards. 

Passing particle orbits in stellarators are complicated by the fact that the magnetic field will in general not be periodic. For a stellarator, we in principle need to calculate the orbit along the entire field line in the domain under consideration. Stellarators also allow for one-sided bounce orbits, which are analogous to hyperbolic orbits in celestial mechanics: we do not consider those here. 

For stellarators, it still can pay to normalize the energy out of the problem and consider characteristic length and time scales. We define the characteristic transit time as
\begin{equation}
    \tau_t = \frac{1}{2 N + 1} \sqrt{\frac{m}{2 \varepsilon}} \int_{-(2 N + 1) \pi}^{(2 N + 1) \pi} \frac{J_{\psi} B \dd{\theta}}{ \sqrt{1 - \lambda B}},
\end{equation}
where $N$ is a non-negative integer. This is the amount of time it takes to traverse a full poloidal turn averaged over $2 N + 1$ turns. We can also define a characteristic length as
\begin{equation}
    L = \frac{1}{2 N + 1} \int_{-(2 N + 1) \pi}^{(2 N + 1) \pi} J_{\psi} B \dd{\theta}.
\end{equation}
The normalized transit time is then
\begin{equation}
    \hat{\tau}_t = \frac{1}{(2 N + 1) L} \int_{-(2 N + 1) \pi}^{(2 N + 1) \pi} \frac{J_{\psi} B \dd{\theta}}{\sqrt{1 - \lambda B}}. 
\end{equation}
The transit frequency can then be defined similarly as in the tokamak case. We then rescale the time coordinate such that
\begin{equation}
    \frac{\dd{\alpha_t}}{\dd{t}} = \Omega_t = \frac{1}{L} \sqrt{\frac{2 \varepsilon}{m}} \hat{\Omega}_t. 
\end{equation}
Then, one solves the equations of motion as in the tokamak case with one's preferred integrator. Note that, unlike in the tokamak case, one also needs to consider both signs of the parallel velocity separately. One can then obtain $\theta(\alpha_t)$. 

\subsubsection{First Order Precession Motion}
We now examine the first order precession motion. The first order guiding center equations of motion satisfy
\begin{align}
    \frac{\dd{\tilde{\psi}}}{\dd{t}} & = \vb{v}_{d} \cdot \nabla \psi, \\
    \frac{\dd{\tilde{\alpha}}}{\dd{t}} & = \vb{v}_{d} \cdot \nabla \alpha. 
\end{align}
Although these equations can be simply integrated over time, some physical intuition is in order. As above, periodicity arguments can be exploited in tokamak geometry. 

When analyzing the drift motion, it is useful to consider how it varies over the transit time. In tokamaks, we thus define the transit average: for any function of $\theta$, we define the transit average as
\begin{equation}
    \left<f(\theta)\right>_t = \frac{1}{\tau_t} \int_{0}^{2 \pi} \frac{\dd{\theta} J_{\psi} B}{|v_{\parallel}|} f(\theta). 
\end{equation}
Equivalently, if we instead know $f$ as a function of $\alpha_t$, then
\begin{equation}
    \left<f(\alpha_t) \right>_t = \int_0^{2 \pi} \frac{\dd{\alpha_t}}{2 \pi} f(\alpha_t). 
\end{equation}
First, we note that in tokamaks
\begin{equation}
    \Omega_{d \psi} = \left<\vb{v} \cdot \nabla \psi \right>_t = 0, 
\end{equation}
meaning that to first order, the particle does not drift away from the flux surface. As for trapped particles, this would be exactly zero if we integrated the guiding center equations exactly in a tokamak. This means that $\tilde{\psi}$ is a periodic function of $\alpha_t$. For simplicity, we just set $\tilde{\psi}(\alpha_t = 0) = 0$ as an initial condition. The solution can be written as
\begin{equation}
    \tilde{\psi}(\alpha_t) = \hat{\psi}(\alpha_t), 
\end{equation}
where $\hat{\psi}$ is a periodic function of $\alpha_t$ and can be computed as
\begin{equation}
    \hat{\psi}(\alpha_t) = \int_0^{\alpha_t} \frac{\dd{\alpha'_t}}{\Omega_t} \vb{v}_{d} \cdot \nabla \psi. 
\end{equation}

Meanwhile, the transit average of $\vb{v}_{d} \cdot \nabla \alpha$ will generally not be zero. Moreover, this quantity is actually not particularly useful to use in tokamaks due to the secular term present in the definition of $\alpha$. We define a different drift frequency as
\begin{equation}
    \Omega_{d \alpha} = \left<\vb{v}_{d} \cdot \nabla \alpha + q' \Omega_t \hat{\psi} + q' \alpha_t \vb{v}_{d} \cdot \nabla \psi \right>_t. 
\end{equation}
Then, one can show
\begin{equation}
    \tilde{\alpha} = \left(\frac{\Omega_{d \alpha}}{\Omega_t} - q' \hat{\psi}(\alpha_t) \right) \alpha_t + \hat{\alpha}(\alpha_t), 
\end{equation}
where $\hat{\alpha}$ is $2 \pi$ periodic. Using the initial condition that $\tilde{\alpha}(\alpha_t = 0) = \hat{\alpha}(\alpha_t = 0)$, we can write the solution as
\begin{equation}
    \hat{\alpha}(\alpha_t) = \int_0^{\alpha_t} \frac{\dd{\alpha'_{t}}}{\Omega_t} \left(\vb{v}_{d} \cdot \nabla \alpha + q' \Omega_t \hat{\psi} + q' \alpha_t \vb{v}_{d} \cdot \nabla \psi - \Omega_{d \alpha} \right). 
\end{equation}
To prove that this is indeed the solution, we need to explicitly write $\alpha = \varphi - q (\theta + \lambda)$, where $\lambda$ is a periodic function that makes $\chi = \theta + \lambda$ straight. Then, 
\begin{equation}
    \vb{v}_{d} \cdot \nabla \alpha = \vb{v}_{d} \cdot \nabla \varphi - q \vb{v}_{d} \cdot \nabla (\theta + \lambda)  - q' \left(\tilde{\theta} + \lambda \right) \vb{v}_{d} \cdot \nabla \psi - q' \alpha_t \vb{v}_{d} \cdot \nabla \psi,
\end{equation}
where the last term on the right-hand side is the secular piece and everything else is periodic. We note that
\begin{equation}
    \frac{\alpha_t}{\Omega_t} \vb{v}_{d} \cdot \nabla \psi = \frac{\dd}{\dd{\alpha_t}} \left(\alpha_t \hat{\psi} \right) - \hat{\psi}. 
\end{equation}
Therefore, we can integrate by parts and find
\begin{equation}
    \int_0^{\alpha_t} \frac{\dd{\alpha_t'}}{\Omega_t} \vb{v}_{d} \cdot \nabla \alpha = - q' \alpha_t \hat{\psi}(\alpha_t) + \int_0^{\alpha_t} \frac{\dd{\alpha_t'}}{\Omega_t} \left(\vb{v}_{d} \cdot \nabla \alpha + q' \Omega_t \tilde{\psi} + q' \alpha_t \vb{v}_{d} \cdot \nabla \psi \right). 
\end{equation}
The terms under the integral on the right-hand side are all periodic, so we can split it up into a monotonically increasing piece and a periodic piece, which is precisely what we did above: 
\begin{equation}
    \int_0^{\alpha_t} \frac{\dd{\alpha_t'}}{\Omega_t} \left(\vb{v}_{d} \cdot \nabla \alpha + q' \Omega_t \tilde{\psi} + q' \alpha_t \vb{v}_{d} \cdot \nabla \psi \right) = \frac{\Omega_{d \alpha}}{\Omega_t} \alpha_t + \hat{\alpha}(\alpha_t). 
\end{equation}
This completes the proof. 

In stellarators, there is strictly speaking no periodic motion to exploit for passing particle trajectories. However, the entire field line is not typically simulated in flux-tube simulations. Instead, one includes only a finite number of poloidal turns \citep{sanchez2021}. In the ballooning representation, one can take advantage of this by periodizing the magnetic field after some number of poloidal turns. Since the magnetic field is then periodic along the field line, the analysis presented in the tokamak-focused sections can be extended to stellarator geometry. In general, it is also necessary to consider a transit average of the radial drift frequency $\Omega_{d \psi}$ in stellarator geometries.

\subsubsection{Useful Integral Identities}
Now that we better understand transit dynamics, we briefly derive some useful expressions. For stellarator geometries, we assume that the geometry has been periodized. We encounter two types of integrals in the main text when integrating over the field line. The first is
\begin{equation}
    \int_{0}^{\theta} \frac{\dd{\theta}' J_{\psi} B}{|v_{\parallel}|} f(\theta') = \tau_t \int_0^{\alpha_t} \frac{\dd{\alpha_t'}}{2 \pi} f\left(\theta(\alpha_t')\right). 
\end{equation}
The second type directly involves the magnetic drift resonance:
\begin{equation}
    \int_{0}^{\theta} \frac{\dd{\theta}' J_{\psi} B}{|v_{\parallel}|} \left(\omega - n \omega_{d} \right), 
\end{equation}
where $\omega$ is the mode frequency, $n$ is the toroidal mode number, and $\omega_{d}$ is defined as
\begin{equation}
    \omega_{d} = \vb{v}_{d} \cdot \left(\nabla \alpha + \theta_0 q' \nabla \psi \right), 
\end{equation}
where $\theta_0$ is the ballooning angle. This can be rewritten as
\begin{equation}
      \int_{0}^{\theta} \frac{\dd{\theta}' J_{\psi} B}{|v_{\parallel}|} \left(\omega - n \omega_{d} \right) = \frac{\omega}{\Omega_t} \alpha_t - n \tilde{\alpha}(\alpha_t) - n \theta_0 q' \tilde{\psi}(\alpha_t). 
\end{equation}
This simplifies to
\begin{equation}
    \int_{0}^{\theta} \frac{\dd{\theta}' J_{\psi} B}{|v_{\parallel}|} \left(\omega - n \omega_{d} \right) = \frac{\alpha_t}{\Omega_t} \left( \omega - n \Omega_{d} - n q' \hat{\psi}(\alpha_t) \right)  - n \tilde{w}_{d}, 
\end{equation}
where
\begin{align}
    \label{eq:drift_passing}
    \Omega_{d} &= \Omega_{d \alpha} - \theta_0 q' \Omega_{d \psi}, \\
    \tilde{w}_{d} & = \hat{\alpha}(\alpha_t) + \theta_0 q' \hat{\psi}(\alpha_t). 
\end{align}
Using action angle variables, we have successfully split these integrals into a secular term and a periodic term. In the main text, it is useful to write 
\begin{equation}
    \int_{0}^{\theta} \frac{\dd{\theta}' J_{\psi} B}{|v_{\parallel}|} \left(\omega - n \omega_{d} \right) = \frac{\alpha_t}{\Omega_t} \left( \omega - n \Omega_{d} \right)  - n \Lambda_d, 
\end{equation}
where $\Lambda_d$ contains all drift deviation terms:
\begin{equation}
    \label{eq:deviations_passing}
    \Lambda_d = \tilde{w}_{d} + \frac{\alpha_t}{\Omega_t} q' \hat{\psi}. 
\end{equation}

\subsection{Trapped Particle Orbits}
\label{sec:trapped}

\subsubsection{Zeroth Order Bounce Motion Motion} 
A trapped particle has parallel velocity equal to zero at two distinct bounce points and to lowest order undergoes periodic motion between those bounce points. The parallel velocity is
\begin{equation}
    v_{\parallel} = \pm \sqrt{\frac{2 \varepsilon}{m}} \sqrt{1 - \lambda B}. 
\end{equation}
For a given $\lambda$, one can determine the bounce points by solving
\begin{equation}
    1 - \lambda B = 0,
\end{equation}
where $B$ is purely a function of the field line following coordinate $\theta$.

Note that in tokamak geometry, we only need to calculate two bounce points; since the magnetic field is periodic, all other pairs of bounce points are equal modulo the period. In a stellarator, the magnetic field is no longer periodic and there will be multiple distinct magnetic wells. Moreover, a trapped particle can traverse over multiple bounce wells at once. In this document, we only consider trapped particle orbits where the particle is confined to a single well, but the equations can be easily extended. 

It is often useful to use the trapped parameter $\kappa$ instead of $\lambda$. We define it as
\begin{equation}
    \kappa^2 = \frac{1 - \lambda B_{\text{min}}}{ 2 \epsilon_B} 
\end{equation}
where
\begin{equation}
    2 \epsilon_B =  1 - \frac{B_{\text{min}}}{B_{\text{max}}}. 
\end{equation}
Here, $B_{\text{min}}$ and $B_{\text{max}}$ are respectively the minimum and maximum of the local magnetic well. Let $\theta_{\text{min}}$ be the location of the minima and $\theta_{\text{max}}$ the location of the smallest nearest maximum. (For a tokamak, $\theta_{\text{max}}$ always comes in pairs that correspond to the same maximum. For a stellarator, there can be two distinct nearest maxima: we pick the smaller one.) The above definition then guarantees that
\begin{align}
    v_{\parallel}(\theta = \theta_{\text{min}}) &= 0 \iff \kappa = 0,\\
    v_{\parallel}(\theta = \theta_{\text{max}}) &= 0 \iff \kappa = 1. 
\end{align}
Therefore, at $\kappa = 0$ the particle is deeply trapped and confined to the bottom of the well. At $\kappa = 1$, the particle is barely trapped in unstable equilibrium at the top of the local well. Moreover,
\begin{align}
    v_{\parallel}^2 &= \frac{2 \varepsilon}{m} \left( 1 - \left(1 - 2 \epsilon_B \kappa^2 \right) \frac{B}{B_{\text{min}}}\right),\\
    v_{\perp}^2 & = \frac{2 \varepsilon}{m} \left(1 - 2 \epsilon_B \kappa^2 \right) \frac{B}{B_{\text{min}}}.
\end{align}
Therefore, all trapped particles satisfy,
\begin{equation}
    0 \le \kappa \le 1.
\end{equation}

The zeroth order bounce motion is periodic, with a bounce period equal to
\begin{equation}
    \tau_b = \oint \dd{\theta} \left(\frac{\dd{\theta}}{\dd{t}}\right)^{-1} = 2 \sqrt{\frac{m}{2 \varepsilon}} \int_{\theta_1}^{\theta_2} \frac{J_{\psi} B \dd{\theta}}{\sqrt{1 - \lambda B}},
\end{equation}
where $\theta_1 < \theta_{\text{min}} < \theta_2$ are the two bounce points. Since we are considering the full forward and then backwards motion, a factor of 2 is included. For tokamaks especially, it is convenient to define a dimensionless bounce time: 
\begin{equation}
    \hat{\tau}_b = \frac{2}{q R_0}  \int_{\theta_1}^{\theta_2} \frac{J_{\psi} B \dd{\theta}}{\sqrt{1 - \lambda B}}. 
\end{equation}
The bounce frequencies (both with and without dimensions) are defined as
\begin{align}
    \Omega_b & = \frac{2 \pi}{\tau_b}, \\
    \hat{\Omega}_b & = \frac{2 \pi}{\hat{\tau}_b}. 
\end{align}

Solving for the bounce motion as a function of time introduces redundancies. If one varies the energy without varying $\lambda$, then one obtains the same motion, just on a different time scale. (This is similar to adjusting the length of a pendulum but keeping the bounce angles the same.) We therefore introduce the action angle variable $\alpha_b$, defined as
\begin{equation}
    \frac{\dd{\alpha}_b}{\dd{t}} = \Omega_b = \frac{1}{q R_0} \sqrt{\frac{2 \varepsilon}{m}} \hat{\Omega}_b. 
\end{equation}
We can then rewrite the second-order equation of motion as
\begin{equation}
    \frac{\dd}{\dd{\alpha_b}} \left(\frac{J_\psi B}{q R_0} \frac{\dd{\theta}}{\dd{\alpha_b}}\right) = - \frac{q R_0}{J_\psi B} \frac{\lambda}{2 \hat{\Omega}_b^2} \frac{\partial B}{\partial \theta}. 
\end{equation}
The modified Hamiltonian is then
\begin{equation}
    \hat{H}(\theta, \hat{p}_{\theta}) = \frac{1}{2} \left(\frac{\hat{p}_{\theta} q R_0}{J_{\psi} B}\right)^2 + \frac{\lambda B}{2 \hat{\Omega}_b^2}.  
\end{equation}
Then, $\theta(\alpha_b)$ is $2 \pi$ periodic. At this point, we can solve for $\alpha_b$ via any ODE integrator. Note that if we wish to make the Hamiltonian separable, that is of the form
\begin{equation}
    H(l, p_l) = \frac{p_l^2}{2} + \frac{f(l)}{2}. 
\end{equation}
once can transform to the normalized arc length $l$ via 
\begin{equation}
    l(\theta) =\frac{1}{q R_0} \int_0^{\theta} \dd{\theta}' J_{\psi} B, 
\end{equation}
in which case
\begin{equation}
    \hat{H} = \frac{p_l^2}{2} + \frac{\lambda B}{2 \hat{\Omega}_b^2}. 
\end{equation}

In terms of initial conditions, it is useful to start at the first bounce point:
\begin{align}
    \theta(\alpha_b = 0) &= \theta_1, \\
    \frac{\dd{\theta}}{\dd{\alpha_b}}(\alpha_b = 0) &= 0. 
\end{align}
This guarantees $\theta(\alpha_b = \pi) = \theta_2$. Due to time-reversibility, $\theta$ is also an even function of $\alpha_2$. (In fact, $\theta(n \pi + \alpha_b) = \theta(n \pi - \alpha_b)$ for all integers $n$.) 

\subsubsection{First Order Drift Motion}
The first order guiding center equations of motion satisfy
\begin{align}
    \frac{\dd{\tilde{\psi}}}{\dd{t}} & = \vb{v}_{d} \cdot \nabla \psi, \\
    \frac{\dd{\tilde{\alpha}}}{\dd{t}} & = \vb{v}_{d} \cdot \nabla \alpha, 
\end{align}
where the magnetic drift velocity $\vb{v}_{d}$ and the quantities $\nabla \theta, \nabla \psi,$ and $\nabla \alpha$ are all evaluated at $\theta(t), \psi_0, \alpha_0$. Although the second set of equations can be simply integrated over time, some physical intuition is in order. 

When analyzing the drift motion, it is useful to consider how it varies over the bounce time. We thus define the bounce average: for any function of $\theta$, we define the bounce average as
\begin{equation}
    \left<f(\theta)\right>_b = \frac{1}{\tau_b} \oint \frac{\dd{\theta} J_{\psi} B}{|v_{\parallel}|} f(\theta), 
\end{equation}
where the closed integral signifies we need to consider forwards and backwards motion. (So for example, $\left<v_{\parallel}\right>_b = 0$). Equivalently, if we instead know the function as a function of $\alpha_b$, then
\begin{equation}
    \left<f(\alpha_b) \right>_b = \int_0^{2 \pi} \frac{\dd{\alpha_b}}{2 \pi} f(\alpha_b). 
\end{equation}
In tokamaks and omnigenous stellarators, we expect
\begin{equation}
     \Omega_{d \psi} = \left<\vb{v}_{d} \cdot \nabla \psi \right>_b = 0, 
\end{equation}
meaning that to first order, the particle does not drift away from the flux surface \citep{helander2014}. Note that if we solved the full guiding center equations of motion in a tokamak or a quasisymmetric stellarator, this would be exact due to the existence of a third invariant (in a tokamak, this would be the canonical toroidal momentum).

If we set the initial condition to be $\tilde{\psi}(\alpha_b = 0) = 0$, then 
\begin{equation}
    \tilde{\psi}(\alpha_b) = \frac{\Omega_{d \psi}}{\Omega_b} \alpha_b + \hat{\psi}(\alpha_b), 
\end{equation}
where, $\hat{\psi}$ is $2 \pi$ periodic and an odd function of $\alpha_b$. (In fact, $\hat{\psi}(n \pi + \alpha_b) = - \hat{\psi}(n \pi - \alpha_b)$ for all integers $n$.) The solution can be written as
\begin{equation}
    \hat{\psi}(\alpha_b) = \int_0^{\alpha_b} \frac{\dd{\alpha_b'}}{\Omega_b} \left(\vb{v}_{d} \cdot \nabla \psi - \Omega_{d \psi}\right). 
\end{equation}

When calculating $\tilde{\alpha}$, we cannot guarantee that the bounce average of $\vb{v}_{d} \cdot \nabla \alpha$ will be zero. This quantity is defined as the bounce-averaged drift frequency and characterizes the drift from field line to field line:
\begin{equation}
    \Omega_{d \alpha} = \left< \vb{v} \cdot \nabla \alpha \right>_b. 
\end{equation}
Therefore, the appropriate solution is
\begin{equation}
    \tilde{\alpha}(\alpha_b) = \frac{\Omega_{d \alpha}}{\Omega_b} \alpha_b + \hat{\alpha}(\alpha_b),
\end{equation}
where
\begin{equation}
    \hat{\alpha}(\alpha_b) = \int_0^{\alpha_b} \frac{\dd{\alpha_b'}}{\Omega_b} \left(\vb{v}_{d} \cdot \nabla \alpha - \Omega_{d \alpha} \right).
\end{equation}
As before, we have chosen $\tilde{\alpha}(\alpha_b = 0)$ as our initial condition, and $\hat{\alpha}$ is a periodic odd function of $\alpha_b$ such that $\hat{\alpha}(n \pi + \alpha_b) = - \hat{\alpha}(n \pi - \alpha_b)$ for all integers $n$.

\subsubsection{Useful Integral Identities}
Now that we better understand bounce dynamics, we briefly derive some useful expressions. We encounter two types of integrals in the main text when integrating over the field line. The first is
\begin{equation}
    2 \int_{\theta_1}^{\theta_2} \frac{\dd{\theta} J_{\psi} B}{|v_{\parallel}|} f(\theta) = \tau_b \left<f(\theta) \right>_b = \tau_b \int_0^{2 \pi} \frac{\dd{\alpha_b}}{ 2 \pi} f\left(\theta(\alpha_b)\right). 
\end{equation}
The second type directly involves the magnetic drift resonance:
\begin{equation}
    \int_{\theta_1}^{\theta} \frac{\dd{\theta}' J_{\psi} B}{|v_{\parallel}|} \left(\omega - n \omega_{d} \right), 
\end{equation}
where $\omega$ is the mode frequency, $n$ is the toroidal mode number, and $\omega_{d}$ is defined as
\begin{equation}
    \omega_{d} = \vb{v}_{d} \cdot \left(\nabla \alpha + \theta_0 q' \nabla \psi \right), 
\end{equation}
where $\theta_0$ is the ballooning angle. This integral can be rewritten as
\begin{equation}
    \int_{\theta_1}^{\theta} \frac{\dd{\theta}' J_{\psi} B}{|v_{\parallel}|} \left(\omega - n \omega_{d} \right) = \frac{\alpha_b}{\Omega_b} \left(\omega - n \Omega_{d} \right) - n \tilde{w}_{d}, 
\end{equation}
where
\begin{align}
\label{eq:drift_trapped}
    \Omega_{d} &= \Omega_{d \alpha} - \theta_0 q' \Omega_{d \psi}, \\
\label{eq:deviations_trapped}
    \tilde{w}_d &= \int_0^{\alpha_b} \dd{\alpha_b'} \frac{\omega_{d} - \Omega_{d \alpha} + \theta_0 q' \Omega_{d \psi}}{\Omega_b} = \hat{\alpha}(\alpha_b) + \theta_0 q' \hat{\psi}(\alpha_b). 
\end{align}
Therefore, when computing the resonant contribution to the gyrokinetic field equations, we are essentially computing the bounce averaged drift frequency as well as the drift deviations along the guiding center orbit. Notice that the equations greatly simplify when working in $\alpha_b$ instead of $\theta$: it pays to use action angle variables! 

Lastly, we note a useful duplication formula for tokamak geometry. Because all bounce wells are identical and spaced $2 \pi$ apart along the field line, we can handle all bounce wells simultaneously. Denote the orbits in the well $2 \pi p$ away from the reference center well with the subscript $p$, where $p$ is any integer. Quantities without a $p$ subscript correspond to the orbits of the reference center well. Then,
\begin{align}
    \Omega_{b, p} & = \Omega_{b}, \\
    \Omega_{d \psi, p} & = 0, \\
    \Omega_{d \alpha, p} & = \Omega_{d \alpha}, \\
    \theta_p(\alpha_b) &= 2 \pi p + \theta(\alpha_b), \\
    \tilde{\psi}_{p}(\alpha_b) & = \hat{\psi}(\alpha_b), \\
    \tilde{\alpha}_{p} ( \alpha_b) & = \frac{\Omega_{d \alpha}}{\Omega_b} \alpha_b + \hat{\alpha} (\alpha_b) - 2 \pi p q' \hat{\psi}(\alpha_b). 
\end{align}
This is because
\begin{align}
    (v_{\parallel})_p & = v_{\parallel}, \\
    (\vu{b} \cdot \nabla B)_p & = \vu{b} \cdot \nabla B, \\
    (\vb{v}_{d} \cdot \nabla \psi)_p & = \vb{v}_{d} \cdot \nabla \psi, \\
    (\vb{v}_{d} \cdot \nabla \alpha)_p &= \vb{v}_{d} \cdot \nabla \alpha - 2 \pi p q' \vb{v}_{d} \cdot \nabla \psi. 
\end{align}
Therefore, we only need to compute the orbits for the $p = 0$ bounce well in tokamak geometry.

\section{Left Eigenvector for the Gyrokinetic Equation}
\label{sec:adjoint}

The electrostatic gyrokinetic equation for each species reads
\begin{equation}
    \left(\omega - v_{\parallel} k_{\parallel} - n \omega_{d s} \right) g_s - \left(\omega - n \omega_{\ast s}\right) \frac{e_s F_{0 s}}{T_s} J_{0 s} \phi = 0. 
\end{equation}
For quasineutrality, we have
\begin{equation}
    \sum_{s} \frac{e_s^2 n_{0 s}}{T_{0 s}} \phi - \int \dd[3]{v} e_s J_{0 s} g_s = 0. 
\end{equation}
For simplicity, we work with two species, ions and electrons. We also set $Z_i = 1$, $n_0 = n_{0i} = n_{0e}$,  $T_{0i} = T_{0e}$, and normalize all quantities.  Let $\bm{\chi}$ be the normalized column vector $ \left(g_i, g_e, \phi \right)^T$. We define our transpose product as
\begin{equation}
    \bm{\chi}_2^T \bm{\chi}_1 = \int \frac{\dd{l}}{B} \left(\int \dd[3]{v} \frac{ g_{i2} g_{i1}}{n_0 F_{0i}} + \frac{n_0 g_{e2}{g_{e1}}}{n_0 F_{0e}} \right) + \int \frac{\dd{l}}{B} \frac{e^2}{T_{0}^2} \phi_2 \phi_1 
\end{equation}
The division by the Maxwellian for the distribution functions ensures that the product converges. The above equations can then be written in block form:
\begin{equation}
    \hat{L} \bm{\chi} = 
    \begin{pmatrix}
        \hat{L}_{ii} &  \hat{L}_{ie} & \hat{L}_{i\phi} \\
        \hat{L}_{ei} &  \hat{L}_{ee} & \hat{L}_{e\phi} \\
        \hat{L}_{\phi i} &  \hat{L}_{\phi e} & \hat{L}_{\phi \phi}
    \end{pmatrix} \begin{pmatrix}
        g_i \\
        g_e \\
        \phi
    \end{pmatrix} = \vb{0}. 
\end{equation}
Here
\begin{align}
    \hat{L}_{ii} &= \omega - v_{\parallel} k_{\parallel} - n \omega_{d i} , \\
    \hat{L}_{ie} &= 0, \\
    \hat{L}_{i \phi} &= - (\omega - n \omega_{\ast i}) F_{0i} J_{0i}, \\
    \hat{L}_{ei} &= 0, \\
    \hat{L}_{ee} &= \omega - v_{\parallel} k_{\parallel} - n \omega_{d e}, \\
    \hat{L}_{e\phi} &= (\omega - n \omega_{\ast e}) F_{0 e} J_{0 e}, \\
    \hat{L}_{\phi i} &= - \int \dd[3]{v} J_{0 i} , \\
    \hat{L}_{\phi e} &= \int \dd[3]{v} J_{0 e} , \\
    \hat{L}_{\phi \phi} &= 2.  
\end{align}
The transpose is
\begin{align}
    \hat{L}_{ii}^T & = \omega + v_{\parallel} k_{\parallel} - n \omega_{d s} , \\
    \hat{L}_{ie}^T &= 0, \\
    \hat{L}_{i \phi}^T  &= -\int \dd[3]{v}  (\omega - n \omega_{\ast i}) J_{0 i}, \\
    \hat{L}_{ei}^T &= 0, \\
    \hat{L}_{ee}^T &= \omega + v_{\parallel} k_{\parallel} - n \omega_{d s}, \\
    \hat{L}_{e\phi}^T &= \int \dd[3]{v} (\omega - n \omega_{\ast e}) J_{0 e}, \\
    \hat{L}_{\phi i}^T &= - J_{0 i} F_{0 i}, \\
    \hat{L}_{\phi e}^T &= J_{0 e} F_{0 e}, \\
    \hat{L}_{\phi \phi}^T &= 2.  
\end{align}
Note the sign change in $v_{\parallel}$. Given adjoint fields $h_i, h_e, \psi$, the adjoint equation is then
\begin{align}
    \left(\omega + v_{\parallel} k_{\parallel} - n \omega_{d i} \right) h_i - J_{0 i} F_{0 i} \psi &= 0, \\
    \left(\omega + v_{\parallel} k_{\parallel} - n \omega_{d e} \right) h_e - J_{0 e} F_{0 e} \psi &= 0, \\
    2 \psi - \int \dd[3]{v} \left(\omega - n \omega_{\ast i} \right) J_{0 i} h_i + \int \dd[3]{v} \left(\omega - n \omega_{\ast e} \right) J_{0 e} h_e & = 0. 
\end{align}
Notice that the equations are markedly different, so the full system is not symmetric. We can, however, make a connection between the old system and the new system. Assume that the fields $(g_i, g_e, \phi)$ solve the old system, and let
\begin{align}
    h_i(z, v_{\parallel}, \mu) &= \frac{1}{\omega - n \omega_{\ast i}} g_i(z, -v_{\parallel}, \mu), \\
    h_e(z, v_{\parallel}, \mu) &= \frac{1}{\omega - n \omega_{\ast e}}  g_e(z, -v_{\parallel}, \mu), \\
    \psi & = \phi. 
\end{align}
Plugging these in (and changing sign of $v_{\parallel} \to - v_{\parallel}$), we obtain
\begin{align}
    \left(\omega + v_{\parallel} k_{\parallel} - n \omega_{d i} \right) h_i - J_{0i} F_{0i} \psi  & \to \left(\omega - v_{\parallel} k_{\parallel} - n \omega_{d i}\right) \frac{g_i}{\omega - n \omega_{\ast i}} - J_{0 i} F_{0 i} \phi, \\
     \left(\omega + v_{\parallel} k_{\parallel} - n \omega_{d e} \right) h_e - J_{0 e} F_{0 e} \psi & \to \left(\omega - v_{\parallel} k_{\parallel} - n \omega_{d e}\right) \frac{g_e}{\omega - n \omega_{\ast e}} - J_{0 e} F_{0 e} \phi
\end{align}
and
\begin{equation}
     2 \psi - \int \dd[3]{v} \left(\omega - n \omega_{\ast i} \right) J_{0i} h_i + \int \dd[3]{v} \left(\omega - n \omega_{\ast e} \right) J_{0e} h_e \to 2 \phi - \int \dd[3]{v} J_{0 i} g_i + \int \dd[3]{v} J_{0 e} g_e. 
\end{equation}
Note that $k_{\parallel}$ commutes with $\omega - n \omega_{\ast}$. Given that $(g_i, g_e, \phi)$ solve the old system, each of the above equations is zero. (The quasineutrality equation is obvious, for the other two just multiply by $\omega - n \omega_{\ast}$.) Therefore, the above guess solves the adjoint system and is thus the left eigenvector of the system. 

\clearpage

\nocite{*}
\bibliographystyle{jpp}
\bibliography{variational_paper}

\end{document}